\documentclass{aa}  

\usepackage[switch]{lineno}
\usepackage{graphicx}
\usepackage{natbib}

\usepackage{supertabular}

\usepackage{txfonts}

\usepackage[]{hyperref}

\usepackage[usenames]{color}

\newcommand{\gdor}{$\gamma$~Dor }
\newcommand{\Kepler}{\textit{Kepler} }

\newcommand{\sap}{LC$_{\rm SAP}$}
\newcommand{\pdc}{LC$_{\rm PDC}$}
\newcommand{\pap}{LC$_\textup{\smash{Pápics}}$}

\usepackage{amssymb}
\usepackage{mathrsfs}

\begin{document} 


   \title{Towards a systematic treatment of observational uncertainties in forward asteroseismic modelling of gravity-mode pulsators}


   \titlerunning{Observational uncertainties in forward asteroseismic modelling of gravity modes}
   
   \author{Dominic M. Bowman 
          \and
          Mathias Michielsen 
          }

    \institute{Institute of Astronomy, KU Leuven, Celestijnenlaan 200D, B-3001 Leuven, Belgium \\
              \email{dominic.bowman@kuleuven.be} 
          }

   \date{Received 6 July 2021 / accepted 22 Sept 2021}

  
 
  \abstract
   {In asteroseismology the pulsation mode frequencies of a star are the fundamental data that are compared to theoretical predictions to determine a star's interior physics. Recent significant advances in the numerical, theoretical and statistical asteroseismic methods applied to main sequence stars with convective cores have renewed the interest in investigating the propagation of observational uncertainties within a forward asteroseismic modelling framework.}
   {We aim to quantify the impact of various choices made throughout the observational aspects of extracting pulsation mode frequencies in main sequence stars with gravity modes.}
   {We use a well-studied benchmark slowly pulsating B star, KIC~7760680, to investigate the sensitivity of forward asteroseismic modelling to various sources of observational uncertainty that affect the precision of the input pulsation mode frequencies.}
   {We quantify the impact of the propagation of the observational uncertainties involved in forward asteroseismic modelling. We find that one of the largest sources of uncertainty in our benchmark star is in the manual building of period spacing patterns, such that the inclusion of a potentially ambiguous pulsation mode frequency may yield differences in model parameters of up to 10\% for mass and age depending on the radial order of the mode.}
   {We conclude that future asteroseismic studies of main sequence stars with a convective core should quantify and include observational uncertainties introduced by the light curve extraction, iterative pre-whitening and the building of period spacing patterns, as these propagate into the final modelling results.}


   \maketitle


\section{Introduction}
\label{section: intro}

In recent decades, asteroseismology has become an established method to accurately probe the interior physics of stars from quantitative comparison of their observed pulsation frequencies with those predicted by theoretical models of stellar structure and evolution \citep{Chaplin2013c, Hekker2017a, Bowman2020c, Aerts2021a}. Across the Hertzsprung--Russell (HR) diagram, stars are able to excite a variety of different types of pulsation modes, which are standing waves with properties defined by stellar structure. Stellar pulsations are typically grouped into pressure (p) modes and gravity (g) modes based on their dominant restoring forces being the pressure force and buoyancy, respectively \citep{ASTERO_BOOK}. However, for rotating stars, the Coriolis force is also important, which gives rise to gravito-inertial modes \citep{Aerts2019b}. Stars can be approximated as spheres, so stellar pulsations are typically described using spherical harmonics. The radial order, $n$, defines the radial geometry of a pulsation mode, and the angular degree, $\ell$, and the azimuthal order, $m$, together define the surface geometry. Identified observed pulsation mode frequencies in terms of $\{n, \ell, m\}$ are then quantitatively compared to a grid of predicted frequencies calculated from models via forward asteroseismic modelling \citep{Aerts2021a}.

In main sequence late-B stars, which are born with a convective core (i.e. $3 \lesssim M \lesssim 8$~M$_{\odot}$), an opacity mechanism operating in the Z~bump is efficient in exciting high radial order (asymptotic) g~modes with periods of order days \citep{Dziembowski1993f, Townsend2005e, Walczak2015}, which defines the group of slowly pulsating B (SPB) stars \citep{Waelkens1991c}. Such a coherent pulsation excitation mechanism is predicted to produce pulsation modes with essentially infinite lifetimes from the observer's perspective. The modes are observed as periodic variations in a star's surface flux and can be extracted by means of Fourier analysis \citep{ASTERO_BOOK}. In reality, however, the majority of coherent pulsators are not observed to have perfectly periodic pulsation modes (see e.g. \citealt{Bowman2016a, VanBeeck2021a*}), which introduces additional uncertainties in the practical analysis of a pulsating star's light curve.

In the last decade significant advances in our understanding of main sequence B-type stars have been made because of asteroseismology. A veritable revolution in the amount and quality of time series data assembled by the CoRoT \citep{Auvergne2009}, Kepler/K2 \citep{Borucki2010, Koch2010, Howell2014}, BRITE \citep{Weiss2014, Weiss2021a} and now TESS \citep{Ricker2015} space missions have provided the necessary continuous, high-precision and long-term data for detailed asteroseismic modelling studies \citep{Aerts2015a, Bowman2020c}. Light curves from space missions have driven increasingly more sophisticated g-mode modelling methods to be developed, which have tested the impact of rotation, interior mixing, convection and opacity tables on the resultant stellar pulsation frequencies \citep{Moravveji2015b, Moravveji2016b, Buysschaert2018c, Szewczuk2018a, Wu_T_2019a, Mombarg2019a, Szewczuk2021a, Pedersen2021a, Michielsen2021a}. Recently, \citet{Aerts2018b} introduced the Mahalanobis distance (MD) within a g-mode forward asteroseismic modelling framework to appropriately deal with the strong correlations and degeneracies among model parameters, because it takes theoretical uncertainties into account. Such a modelling setup has been recently applied to 26 SPB stars observed by the \Kepler mission and revealed a large range in the amount of envelope mixing \citep{Pedersen2021a}. Furthermore, the differences and advantages of using the MD compared to the $\chi^2$ merit function in the context of g-mode asteroseismology have been extensively demonstrated by \citet{Michielsen2021a}.

From the observational perspective, the methods of how one extracts pulsation frequencies and incorporates the various sources of observational uncertainty at the different stages of analysing time series photometry sometimes yields significantly different results (see e.g. \citealt{Handler2019a}). This can then propagate into the process of mode identification in g-mode pulsators relying on the morphology of period spacing patterns. Such patterns are built to satisfy the requirement of consecutive radial order modes of the same angular degree and azimuthal order from a list of extracted pulsation frequencies, but typically less than 10\% of the total number of extracted pulsation frequencies are used in the resultant period spacing pattern (see e.g. \citealt{Papics2014, Papics2015, Papics2017a, Christophe2018a}). In some cases multiple g-mode period spacing patterns for a star are plausible and it is not clear which of them are optimal for modelling (see e.g. \citealt{Szewczuk2018a, Szewczuk2021a, Pedersen2021a}).

In this paper, motivated by the recent modelling study of \citet{Michielsen2021a} of the SPB star KIC~7760680, we explore the implications of various choices within the process of extracting its pulsation mode frequencies and g-mode period spacing pattern. In section~\ref{section: method} we justify our choice of KIC~7760680, and in section~\ref{section: steps} we describe the different approaches to extracting and analysing light curves and in the building of g-mode period spacing patterns. In section~\ref{section: results} we demonstrate the impact of these different approaches on the resultant best-fit model parameters when performing forward asteroseismic modelling, and we conclude in section~\ref{section: conclusions}.


\section{Benchmark star: KIC~7760680}
\label{section: method}

Within forward asteroseismic modelling of main-sequence B-type stars pulsating in high-radial order g~modes (i.e. SPB stars), the requirement is to extract and identify pulsation mode frequencies from a time series (e.g. light curve). Such stars have typically been found to exhibit prograde ($m=1$) or zonal ($m=0$) dipole ($\ell = 1$) modes spanning from a few to tens of consecutive radial order modes. Although a non-negligible fraction of these stars have dipole retrograde ($m=-1$) patterns, or those formed of quadrupole ($\ell = 2$) modes \citep{Degroote2010a, Papics2011, Papics2014, Papics2015, Papics2017a, Christophe2018a, Szewczuk2021a, Pedersen2021a}. The choices of how to extract and identify the pulsation mode frequencies, however, can differ in the literature.

We note it is also necessary to include additional constraints in forward asteroseismic modelling to delimit the large parameter space within the HR~diagram, such as spectroscopic estimates of the effective temperature ($T_{\rm eff}$), metallicity ($[{\rm M} / {\rm H}]$), and luminosity ($L$) or surface gravity ($\log\,g$). Not only does this drastically reduce the number of stellar structure models needed, but the best asteroseismic model is typically selected from within the spectroscopic error box in the HR~diagram (see e.g. \citealt{Moravveji2015b, Moravveji2016b, Michielsen2021a}). In this paper, we do not investigate the propagation of uncertainties arising from the choices in setup and determination of spectroscopic parameters (see e.g. \citealt{Tkachenko2020a}). The fractional uncertainties of pulsation mode frequencies are typically several orders of magnitude smaller compared to spectroscopic parameters in pulsating early-type stars, and are used to fine-tune the physics of the best model.

To perform a quantitative investigation of the impact of choices in the extraction of g-mode period spacing patterns, we selected the SPB star KIC~7760680 because:
\begin{itemize}
\item it was observed continuously during the 4-yr \Kepler mission, hence its long light curve has high resolving power for extracting individual frequencies;
\item it is a bright star ($V = 10.3$~mag) and thus its light curve has high photometric precision;
\item it has precise spectroscopic parameters available and the star is known to be single \citep{Papics2015, Gebruers2021a}; and
\item it has a long series of consecutive radial order g~modes with high pulsation mode amplitudes.
\end{itemize}

\noindent Moreover, KIC~7760680 was recently the subject of detailed modelling efforts \citep{Moravveji2016b, Pedersen2021a, Michielsen2021a}. Most importantly, these modelling efforts have all implicitly relied on the period spacing pattern extracted in the discovery paper by \citet{Papics2015}. The known spectroscopic parameters of KIC~7760680 are given in Table~\ref{table: spectra}.

\begin{table}
\caption{Spectroscopic parameters for KIC~7760680 determined by \citet{Papics2015}.} 
\begin{center}
\resizebox{0.99\columnwidth}{!}{
\begin{tabular}{c c c c c}
\hline \hline
$T_{\rm eff}$	&	$\log\,g$	&	$v_{\rm micro}$		&	$v\,\sin\,i$		&	$[{\rm M} / {\rm H}]$	\\
(K)	&	(dex)		&	(km\,s$^{-1}$)	&	(km\,s$^{-1}$)	&	(dex)		\\
\hline
$11650 \pm 210$	&	$3.97 \pm 0.08$ 	&	$0.0^{+0.6}_{-0.0}$	&	$61.5 \pm 0.5$	&	$0.14 \pm 0.09$	\\
\hline
\end{tabular} }
\end{center}
\label{table: spectra}
\end{table}


\section{From a telescope to a period spacing pattern}
\label{section: steps}

\begin{figure}
\centering
\includegraphics[width=0.99\columnwidth]{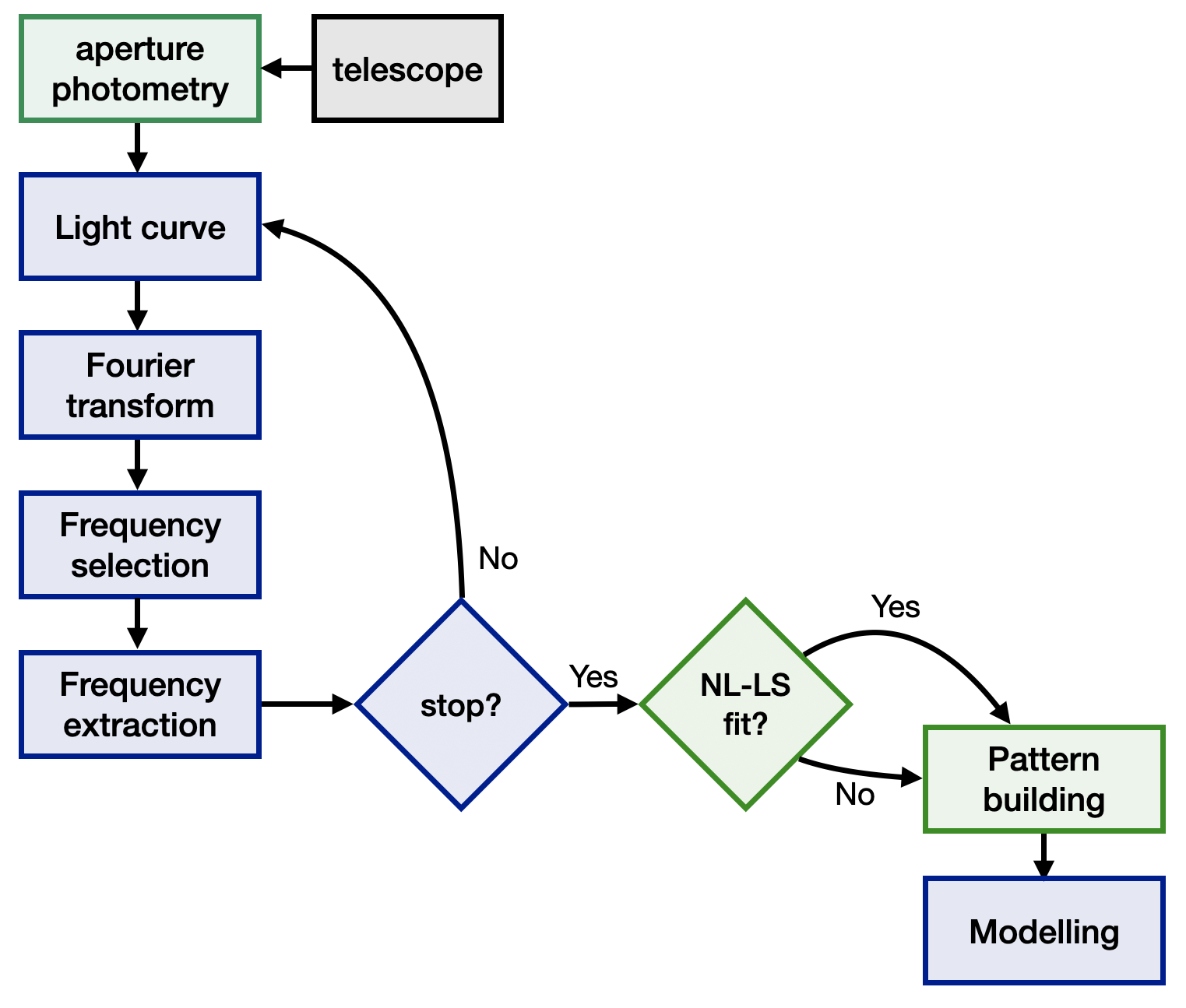}
\caption{Schematic overview of the various steps involved in observational aspects prior to forward asteroseismic modelling of g~modes in main-sequence stars. The steps denoted in green are those we assess the impact of in this work.}
\label{figure: flowchart}
\end{figure}

 In Fig.~\ref{figure: flowchart} we illustrate the various steps that are involved up until forward asteroseismic modelling of g-mode pulsators, but the implementation of these steps varies within the literature. In the following subsections we describe each step in more detail and highlight the choices that are typically made.


	\subsection{Light curve extraction}
	\label{subsection: extraction}
	
	For targets from the nominal \Kepler mission, there are different data products publicly available via the Mikulski Archive for Space Telescopes (MAST)\footnote{MAST website: \url{https://archive.stsci.edu/}}. Each \Kepler target was assigned a postage stamp of pixels to download, such that image reduction and light curve extraction was performed on ground. Two types of light curves from these postage stamps were extracted for each target. The first data product are light curves extracted using simple aperture photometry (SAP), and the second are those from the \Kepler mission pipeline (PDC-SAP; \citealt{Jenkins2010a}). The \Kepler mission pipeline produces light curves that are noise optimised, which means that the selected apertures are relatively small to minimise the contribution of the sky background. However, smaller apertures are more prone to target drift and thermal recovery of the \Kepler satellite after pointing, which can produce long-period trends in the extracted time series. 
	
	However, few studies have investigated the differences in g-mode pulsation frequencies obtained from different light curve extractions. \citet{Tkachenko2013b} identified instrumental frequencies in a sample of \gdor stars observed by \Kepler by comparing the analysis of PDC-SAP and customised light curves, and demonstrated the importance of creating optimised light curves for a subsequent frequency analysis. The customised light curve and period spacing pattern extractions for the SPB star KIC~7760680 by \citet{Papics2015} were later used by \citet{Moravveji2016b}, \citet{Pedersen2021a} and \citet{Michielsen2021a} for modelling purposes. Yet, these studies assumed the observations to be free of any systematic observational uncertainties and biases in how the light curve and pattern were extracted. Furthermore, the method of \citet{Papics2015} extracted and estimated the uncertainties of pulsation mode frequencies assuming they were all completely independent, which is an idealistic scenario. 
	
\begin{figure}
\centering
\includegraphics[width=0.99\columnwidth]{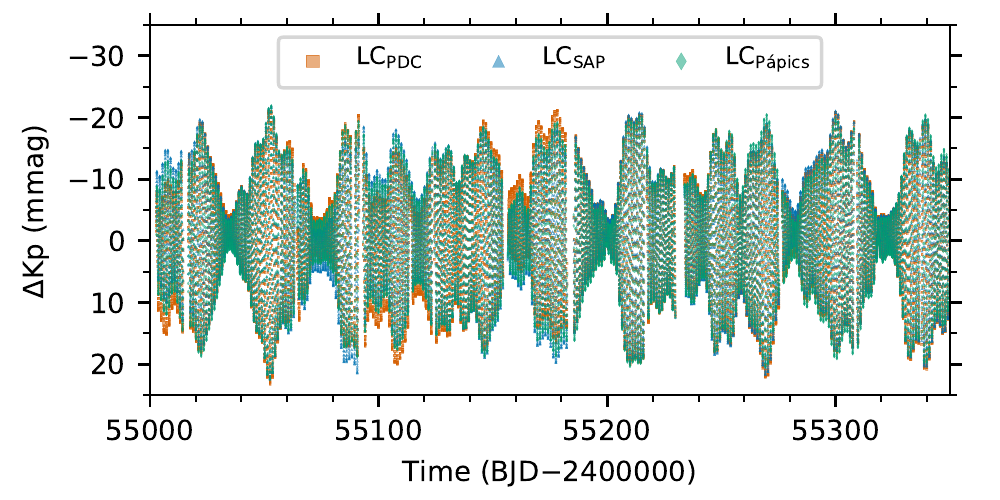}
\caption{1-yr excerpts of the \sap{}, \pdc{} and \pap{} light curves of KIC~7760680, in which the high-amplitude pulsations can be seen.}
\label{figure: light curves}
\end{figure}

	We show the SAP and PDC-SAP light curves from data release (DR) 25 of the \Kepler mission pipeline and the \citet{Papics2015} light curve of KIC~7760680 in Fig.~\ref{figure: light curves}. We refer to these as the \sap{}, \pdc{}, and \pap{} light curves, and consider them as three representative light curves of KIC~7760680 that one may consider to analyse for the purpose of forward asteroseismic modelling. The \sap{} and \pdc{} light curves were downloaded from MAST and converted to have flux units of magnitudes. The customised \pap{} light curve was extracted directly from the \Kepler postage stamps using a larger aperture mask and underwent additional de-trending using a low-order polynomial \citep{Papics2015}. For consistency, we convert the \pap{} light curve to have flux units of magnitudes.  All three light curves are normalised to have a median flux of zero.
	
\begin{figure}
\centering
\includegraphics[width=0.99\columnwidth]{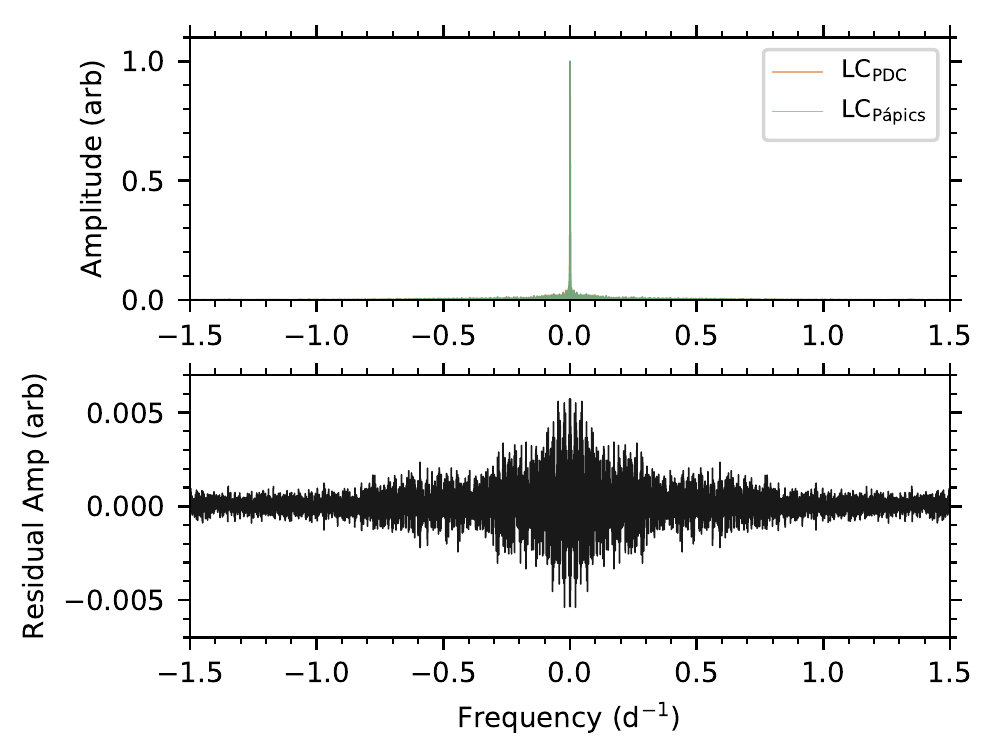}
\caption{Spectral windows of the \pdc{} and \pap{} light curves of KIC~7760680. The difference between the spectral windows of these light curves is caused by difference in the number of data points (i.e. duty cycle) in the \pap{} light curve given that the start and end time stamp for all light curves are the same.}
\label{figure: windows}
\end{figure}

	The \sap{}, \pdc{} and \pap{} light curves of KIC~7760680 shown in Fig.~\ref{figure: light curves} look nearly identical. The statistical properties including the total number of data points, the total time span, the maximum and minimum fluxes, and the standard deviation of the fluxes are given in Table~\ref{table: light curve stats}. The start and end time stamps of all three light curves (i.e. BJD~2454953.538827 and 2456424.001160) are exactly the same yielding a total time span of 1470.462~d. However, the number of data points in the \pap{} light curve is slightly larger compared to the \sap{} and \pdc{} light curves. The differences in the minimum, maximum and standard deviation of the flux values of each light curves are because of the different data reduction steps taken in extracting each light curve. We plot the spectral window of the \pdc{} and \pap{} light curves in Fig.~\ref{figure: windows}, but not of the \sap{} light curve because it is identical to that of the \pdc{} light curve. As shown in Fig.~\ref{figure: windows}, the difference in the spectral windows of the \pdc{} and \pap{} light curves are notable at the $\sim$0.5\% level in the $\pm 1$~d$^{-1}$ frequency range. 
	
\begin{table}
\caption{Statistical properties of the complete 4-yr \sap{}, \pdc{} and \pap{} light curves of KIC~7760680.} 
\begin{center}
\begin{tabular}{l c c c c c}
\hline \hline
\multicolumn{1}{c}{} & \multicolumn{1}{c}{$N_{\rm data}$} & \multicolumn{1}{c}{max(flux)} & \multicolumn{1}{c}{min(flux)} & \multicolumn{1}{c}{std(flux)} \\
\multicolumn{1}{c}{} & \multicolumn{1}{c}{} & \multicolumn{1}{c}{(mmag)} & \multicolumn{1}{c}{(mmag)} & \multicolumn{1}{c}{(mmag)} \\
\hline
\sap{}		&	$65263$	&	$24.7056$	&	$-26.0299$	&	$9.2244$	\\
\pdc{}		&	$65263$	&	$27.0814$	&	$-23.9249$	&	$9.1888$	\\
\pap{}		&	$65898$	&	$24.9922$	&	$-24.9852$	&	$9.1340$	\\
\hline
\hline
\end{tabular}
\end{center}
\label{table: light curve stats}
\end{table}

	\subsection{Iterative pre-whitening}
	\label{subsection: iterative pre-whitening}
	
	After a light curve has been extracted, the next step is to extract significant and independent pulsation mode frequencies using iterative pre-whitening. The premise of iterative pre-whitening in asteroseismology of coherent pulsators assumes that each individual pulsation mode within a time series can be represented by a (co)sinusoid with a constant amplitude, $A$, frequency, $\nu$, and phase, $\phi$. Thus, the Fourier decomposition of a time series reveals the individual pulsation mode frequencies. It is more common within this framework to extract pulsation mode frequencies iteratively by sequentially extracting them in order of decreasing amplitude, and optimising their parameters using a least-squares fit to the light curve with the equation:
	\begin{equation}
	\Delta\,m = \sum_{i}^{N} A_i\,\cos\left(2\pi\nu_i\left(t-t_0\right)+\phi_i\right) ~ ,
	\label{equation: cosine}
	\end{equation}
	\noindent where $t$ is the time with respect to a zero-point $t_0$. In this way, optimising $N$ cosinusoids (each with a frequency, $\nu_i$, amplitude, $A_i$, and phase, $\phi_i$) sequentially using Eqn.~(\ref{equation: cosine}) allows one to subtract the least-squares model from the light curve, and find the next frequency using an iterative approach. This methodology is the backbone of the commonly-used {\sc Period04} software package designed for this specific purpose \citep{Lenz2004b, Lenz2005}. 
	
	We note that there is some variance in the literature on how the Fourier transform of the input time series is calculated. For example, {\sc Period04} and some studies utilise the Discrete Fourier Transform (DFT) for unevenly-sampled data \citep{Deeming1975, Kurtz1985b} to directly calculate a frequency spectrum (see e.g. \citealt{Lenz2004b, Kurtz2015b, Bowman_BOOK}). Whereas other codes prefer (a variant of) the Lomb-Scargle algorithm \citep{Scargle1982a, Press1989a} for calculating the periodogram of a light curve (see e.g. \citealt{Degroote2009a, Antoci2019a, VanBeeck2021a*}). In this work we utilize the former method, such that our results are fully consistent and verifiable with the {\sc Period04} software package. 
	
	\subsubsection{Fitting of sinusoids with least-squares}
	
	It has recently been shown that SPB stars exhibit amplitude and frequency modulation and long-period beating of closely-spaced frequencies \citep{VanBeeck2021a*}. Furthermore, SPB stars also exhibit stochastically-excited gravity waves, which lead to damped modes with finite lifetimes \citep{Bowman2019a, Bowman2019b, Neiner2020b}. Therefore not all signals extracted using iterative pre-whitening represent independent and perfectly periodic pulsation modes. In all implementations of iterative pre-whitening, the light curve is decomposed into harmonic terms, which are then interpreted as either pulsation modes or not (see e.g. \citealt{Poretti2009, Chapellier2011}).
	
	Some studies that utilise iterative pre-whitening perform a least-squares fit in which the frequency is kept fixed to the Fourier frequency taken from the amplitude spectrum. In this implementation the assumption is that all independent frequencies are well described by the corresponding Fourier frequency and have uncorrelated uncertainties. This can lead to a large number of extracted frequencies, especially when combined with a non-conservative significance criterion (see e.g. \citealt{Papics2012a, Balona2014b}).
	
	A multi-cosinusoid non-linear least-squares fit using Eqn.~(\ref{equation: cosine}), in which $\nu_i$, $A_i$ and $\phi_i$ are all free parameters both during and at the end of iterative pre-whitening, has the advantage of taking the covariance of pulsation mode frequencies, amplitudes and phases into account and allows for the calculation of correlated uncertainties (see section~\ref{subsubsection: frequency comparison}). This is typically done as a second step such that the fixed-frequency least-squares solution is used as input to the multi-cosinusoid non-linear least-squares fit, which outputs optimised frequencies, amplitudes and phases and uncertainties determined from the error matrix \citep{Lenz2005, Kurtz2015b, Bowman_BOOK}. Frequency and phase are highly correlated in a cosinusoid model, and a non-linear least-squares fit of hundreds of cosinusoids can become an unwieldy numerical problem and lead to ill-conditioned covariance and error matrices. Thus, an appropriate compromise is to perform a multi-cosinusoid non-linear least squares fit after all spurious and instrumental frequencies have been removed \citep{ASTERO_BOOK, Bowman_BOOK, VanBeeck2021a*}.
	
	\subsubsection{Optimum significance criterion}
	
	The most commonly used criterion is the empirically-derived signal-to-noise (S/N) criterion of \citet{Breger1993b}, which defines significant frequencies to have an amplitude S/N~$\geq4$ in the frequency spectrum. However, it is based on ground-based data of p-mode pulsators, such that its applicability to space photometry of g-mode pulsators is questionable. Some studies advocate for a higher threshold for space telescope data because there is an increased likelihood of classifying instrumental frequencies as astrophysical \citep{Kuschnig1997c, Baran2015b, Zong2016a}. Whereas others advocate for S/N~$\geq3$ when extracting g~modes based on knowledge from the expert user (e.g. \citealt{Li_G_2019a}). Statistical investigations of light curves from space telescopes using $p$-values over-estimate the number significant frequencies compared to a S/N criterion (e.g. \citealt{Degroote2009a, Blomme2011b, Papics2012a}). Moreover, the results of five different iterative pre-whitening strategies applied to 38 \Kepler SPB stars revealed large differences in the number of significant frequencies \citep{VanBeeck2021a*}. 
	
\begin{figure}
	\centering
	\includegraphics[width=0.99\columnwidth]{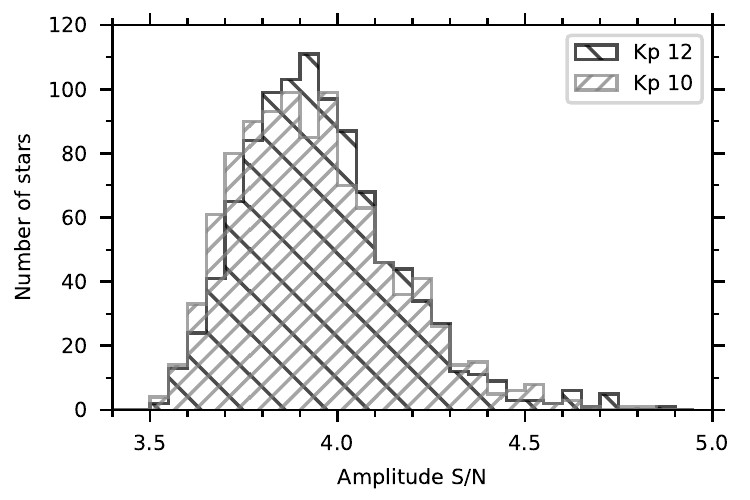}
	\caption{Histogram showing the signal-to-noise ratio (S/N) of the highest amplitude frequency extracted from 1000 synthetic light curves containing only white noise consistent with \Kepler targets of 10 and 12~mag.}
	\label{figure: histogram}
	\end{figure}

	To assess the validity of the S/N~$\geq 4$ criterion in g-mode asteroseismology, we create two sets of 1000 synthetic light curves. We use the long-cadence time stamps of KIC~7760680, and calculate white Gaussian noise typical of stars with \Kepler instrumental magnitudes (Kp) of 10 and 12~mag in the two sets. We calculate the amplitude spectra of each light curve up to the Nyquist frequency (i.e. 24.45~d$^{-1}$) and extract the S/N value of the highest amplitude (noise) peak as the ratio of the peak's amplitude to the local noise level using a 1-d$^{-1}$ frequency window in the residual amplitude spectrum. The histograms for the resultant S/N values from the 1000 synthetic light curves of each brightness subset are shown in Fig.~\ref{figure: histogram}. 
		
	The similarities of the two histograms in Fig.~\ref{figure: histogram} demonstrate how the S/N of noise peaks in a \Kepler light curve are not wholly sensitive to the shot noise level as set by the brightness of the star. There are 339 and 359 light curves in the Kp~10 and 12~mag subsets whose tallest (noise) peak satisfies S/N~$\geq 4$, respectively. To reduce the false alarm probability of extracting noise peaks as significant frequencies to 10 in 1000 (i.e. 1\%), one would need to impose a S/N threshold of 4.52 and 4.60 for the Kp~10 and 12~mag sets of synthetic light curves, respectively\footnote{See also similar tests by \citet{Zong2016a} based on short-cadence \Kepler data of pulsating sub-dwarf stars.}. Frequencies with $3.5 <$~S/N~$< 4.0$ extracted from \Kepler photometry are indistinguishable from high-amplitude noise, and those between $4.0 \leq$~S/N~$\leq 4.6$ have a modest chance of being noise peaks. Hence as a conservative significance threshold we advocate using S/N~$\geq 4.6$ within an iterative pre-whitening methodology for \Kepler data. 

	\subsubsection{Comparison of extracted frequencies}
	\label{subsubsection: frequency comparison}
	
\begin{figure}
\centering
\includegraphics[width=0.99\columnwidth]{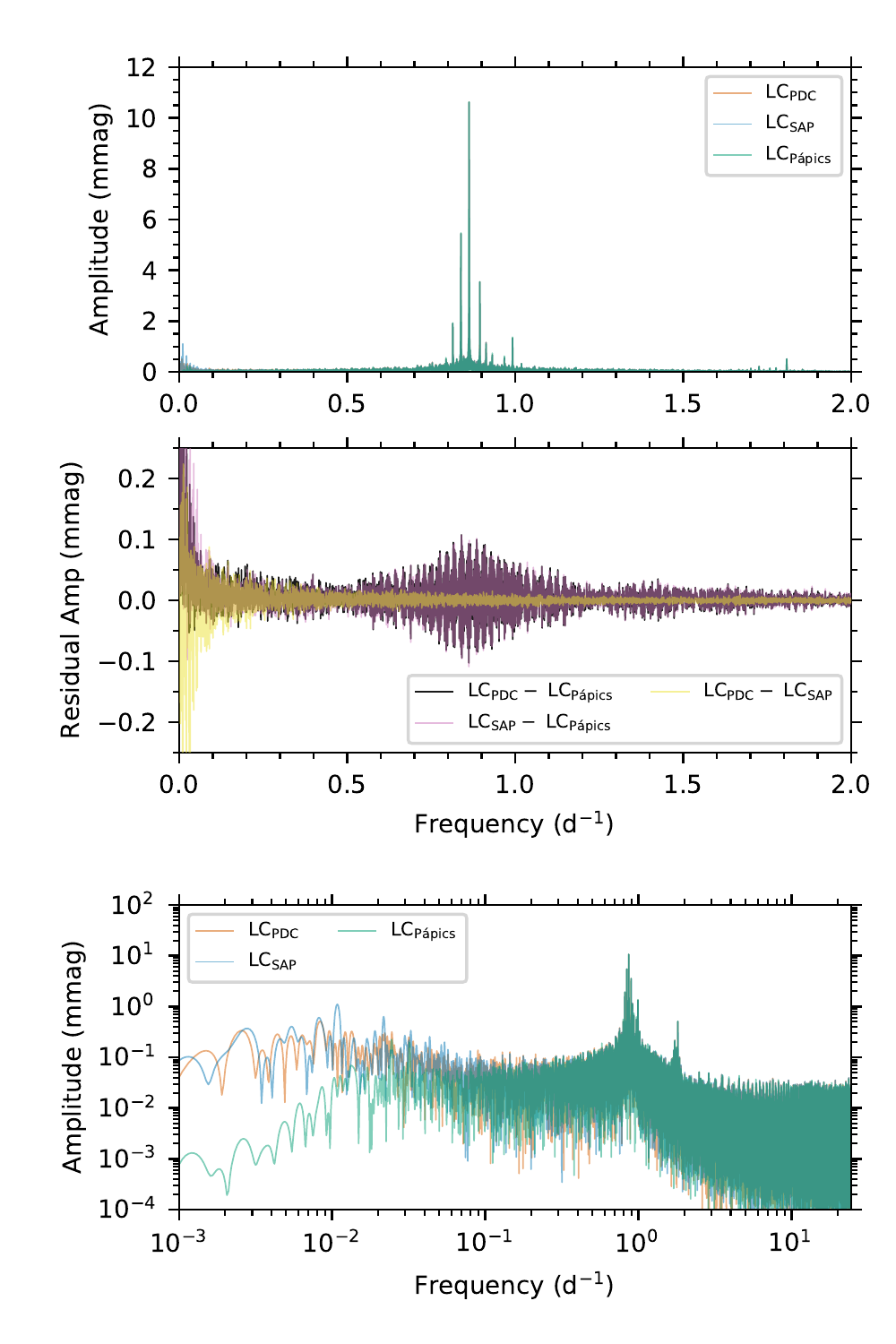}
\caption{Top panel: amplitude spectra of the \sap{}, \pdc{} and \pap{} light curves of KIC~7760680, in which the low frequency g-mode pulsation modes can be seen. Middle panel: pair-wise difference in amplitude spectra shown in the top panel. Bottom panel: logarithmic amplitude spectra up to the \Kepler long cadence Nyquist frequency of 24.4662~d$^{-1}$.}
\label{figure: FT differences}
\end{figure}

	With all this is mind, we calculate the amplitude spectra of the \sap{}, \pdc{}, and \pap{} light curves of KIC~7760680 by means of a DFT \citep{Kurtz1985b}. All three light curves have the same cadence\footnote{Strictly speaking, we mean the same average cadence, given that \Kepler data have a periodic Nyquist frequency (see \citealt{Murphy2013a}).} and total time span, so all have the same Nyquist frequency (24.4662~d$^{-1}$) and Rayleigh resolving power (0.00068~d$^{-1}$). We use a frequency oversampling factor of 20 when calculating the amplitude spectra. The three amplitude spectra of the \sap{}, \pdc{} and \pap{} light curves are shown in Fig.~\ref{figure: FT differences}. We plot the pair-wise difference of each amplitude spectrum for the low-frequency regime in Fig.~\ref{figure: FT differences}. By design, the \Kepler pipeline aims to remove much of the long-period systematics from the \sap{} light curve when producing the \pdc{} light curve, as shown in Fig.~\ref{figure: FT differences}. There is also a dramatic reduction in the low-frequency noise (i.e. $\nu < 0.1$~d$^{-1}$) in the \pap{} light curve, which is primarily the result of the larger pixel mask \citep{Papics2015}. To emphasise this effect, we also include the logarithmic amplitude spectra of all three light curves in Fig.~\ref{figure: FT differences}. The reduction in noise in the high frequency regime (i.e. $\nu > 10$~d$^{-1}$) is also significant, with the shot noise being on average $13\%$ lower in the amplitude spectrum of \pap{} compared to \pdc{} light curve.
		
\begin{figure}
\centering
\includegraphics[width=0.97\columnwidth]{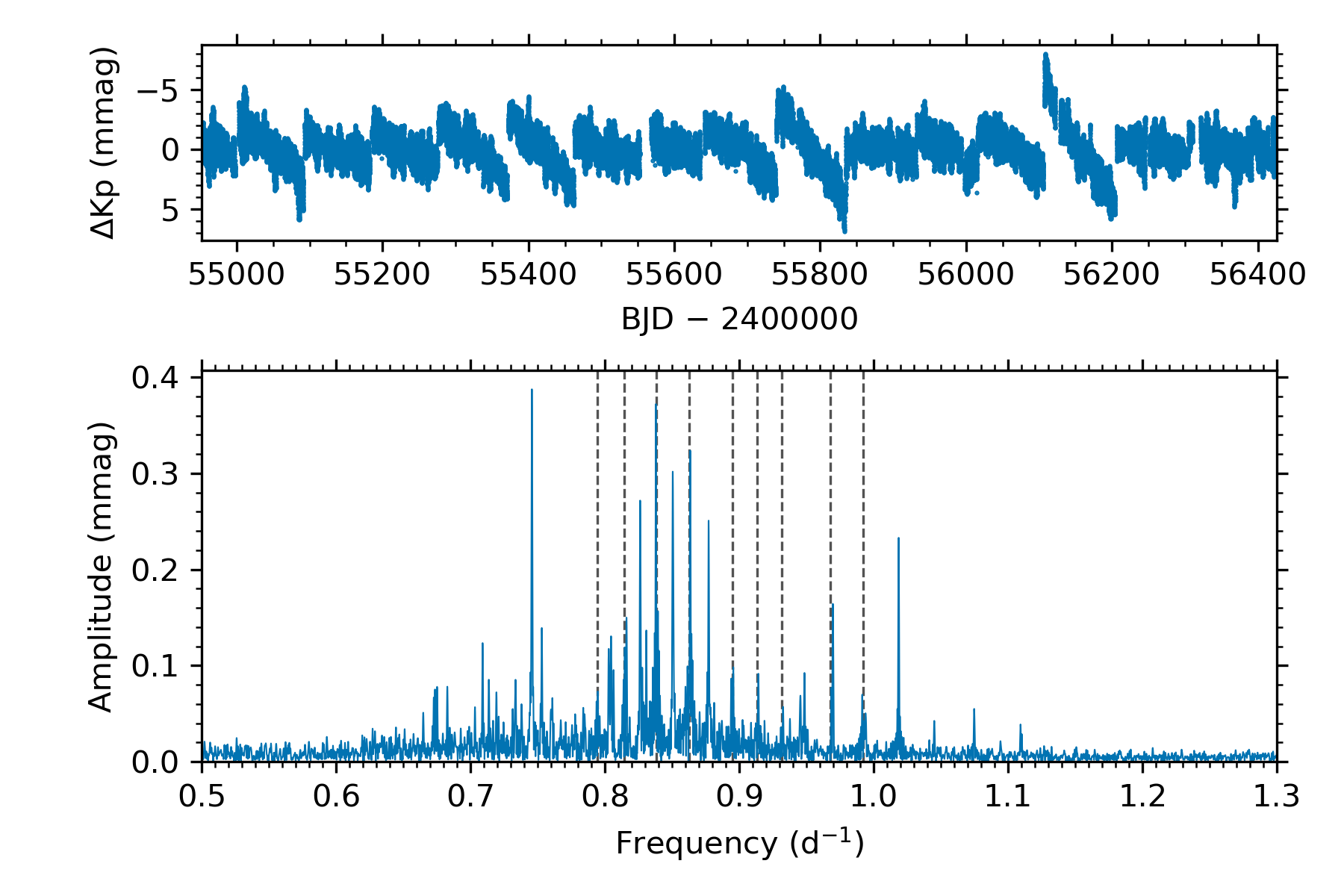}
\includegraphics[width=0.97\columnwidth]{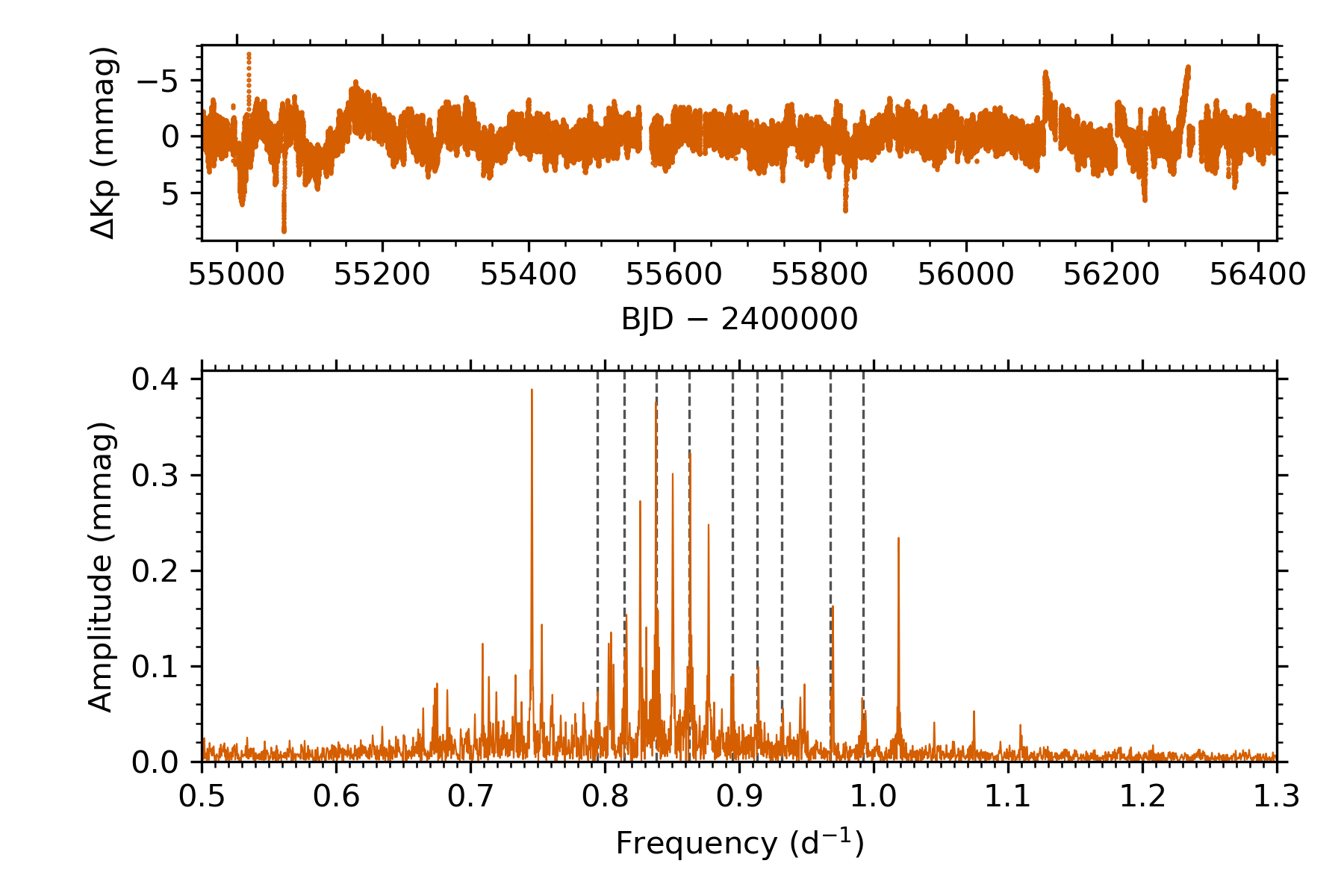}
\includegraphics[width=0.97\columnwidth]{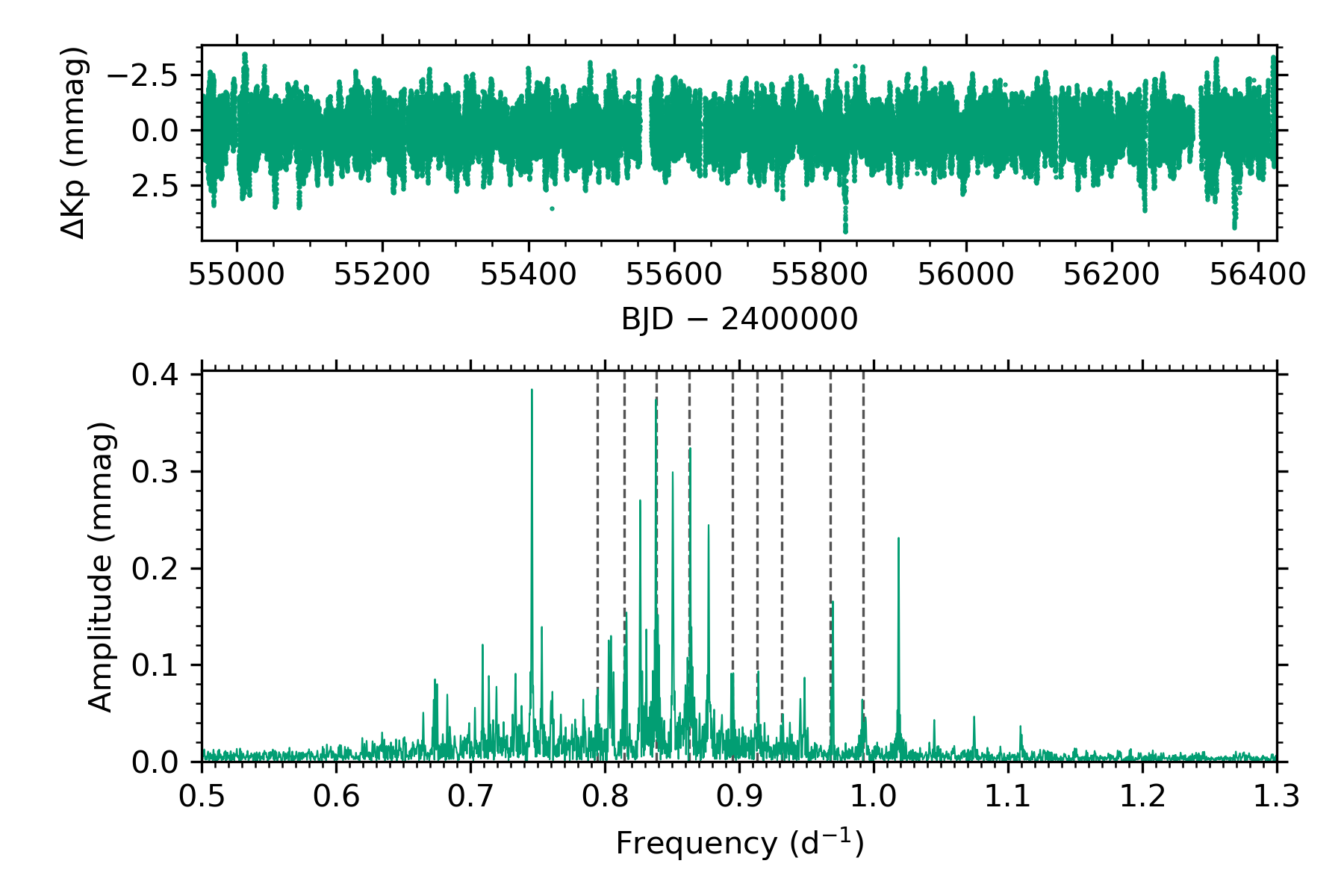}
\caption{Residual light curves and amplitude spectra of the \sap{} (top), \pdc{} (middle) and \pap{} (bottom) light curves of KIC~7760680, after the nine highest-amplitude frequencies have been removed during iterative pre-whitening. The full 4-yr light curves are shown but only a small frequency region containing the $\ell=1$ period spacing pattern is shown in the amplitude spectra for visibility. The nine frequencies extracted from each light curve (shown as dashed vertical lines) are given in Table~\ref{table: single-freq}. Note the difference in ordinate-axis scale.}
\label{figure: iteration 9}
\end{figure}
	
	The differences among the three amplitude spectra in Fig.~\ref{figure: FT differences} are maximal at the location of the g~mode frequencies (i.e. $0.6 \leq \nu \leq 1.1$~d$^{-1}$), which are caused by the interference of the spectral windows of multiple g~modes. We extract the dominant pulsation mode, optimise its parameters using a non-linear least-squares fit to each light curve separately and provide the resultant frequencies, amplitudes and phases in Table~\ref{table: single-freq}. We also provide the S/N value calculated using a window of 1~d$^{-1}$ centred on the extracted frequency to estimate of the local noise level in the residual amplitude spectrum. From a comparison of the dominant pulsation mode (i.e. frequency label $1$ in Table~\ref{table: single-freq}), the resultant frequencies, amplitudes and phases are consistent within $1\sigma$ across all three light curves.

\begin{table*}
\caption{Single-cosinusoid non-linear least-squares solutions for the first nine iterations for the three light curves of KIC~7760680, such that each frequency is extracted and optimised independently. The zero point of the time scale is $t_0 = 2455688.770$~BJD.} 
\begin{center}
\begin{tabular}{l c r r r r}
\hline \hline
\multicolumn{1}{c}{} & \multicolumn{1}{c}{Freq. label} & \multicolumn{1}{c}{Frequency} & \multicolumn{1}{c}{Amplitude} & \multicolumn{1}{c}{Phase} & \multicolumn{1}{c}{S/N} \\
\multicolumn{1}{c}{} & \multicolumn{1}{c}{} & \multicolumn{1}{c}{(d$^{-1}$)} & \multicolumn{1}{c}{(mmag)} & \multicolumn{1}{c}{(rad)} & \multicolumn{1}{c}{} \\
\hline
\sap{}		&	1	&	$0.8628724 \pm 0.0000010$	&	$10.658 \pm 0.030$	&	$0.4982 \pm 0.0028$	&	$141.19$	\\
\pdc{}		&	1	&	$0.8628723 \pm 0.0000010$	&	$10.652 \pm 0.029$	&	$0.4981 \pm 0.0027$	&	$141.11$	\\
\pap{}		&	1	&	$0.8628727 \pm 0.0000010$	&	$10.665 \pm 0.028$	&	$0.4992 \pm 0.0027$	&	$146.34$	\\
\hline 
\pdc{}		&	2	&	$0.8385280 \pm 0.0000014$	&	$5.451 \pm 0.020$	&	$-2.2739 \pm 0.0037$	&	$104.20$	\\
\pdc{}		&	2	&	$0.8385282 \pm 0.0000014$	&	$5.457 \pm 0.020$	&	$-2.2739 \pm 0.0036$	&	$104.27$	\\
\pap{}		&	2	&	$0.8385273 \pm 0.0000013$	&	$5.448 \pm 0.019$	&	$-2.2751 \pm 0.0035$	&	$108.12$	\\
\hline 
\sap{}		&	3	&	$0.8948497 \pm 0.0000015$	&	$3.695 \pm 0.014$	&	$2.2767 \pm 0.0039$	&	$98.81$	\\
\pdc{}		&	3	&	$0.8948486 \pm 0.0000014$	&	$3.699 \pm 0.014$	&	$2.2774 \pm 0.0037$	&	$99.28$	\\
\pap{}		&	3	&	$0.8948487 \pm 0.0000013$	&	$3.693 \pm 0.012$	&	$2.2735 \pm 0.0033$	&	$103.42$	\\
\hline 
\sap{}		&	4	&	$0.8143166 \pm 0.0000024$	&	$1.915 \pm 0.012$	&	$-1.0916 \pm 0.0064$	&	$61.81$	\\
\pdc{}		&	4	&	$0.8143177 \pm 0.0000022$	&	$1.915 \pm 0.011$	&	$-1.0907 \pm 0.0059$	&	$62.12$	\\
\pap{}		&	4	&	$0.8143186 \pm 0.0000019$	&	$1.913 \pm 0.010$	&	$-1.0872 \pm 0.0051$	&	$64.99$	\\
\hline 
\sap{}		&	5	&	$0.9921895 \pm 0.0000027$	&	$1.509 \pm 0.011$	&	$1.6430 \pm 0.0071$	&	$58.97$	\\
\pdc{}		&	5	&	$0.9921889 \pm 0.0000024$	&	$1.519 \pm 0.010$	&	$1.6422 \pm 0.0063$	&	$59.63$	\\
\pap{}		&	5	&	$0.9921894 \pm 0.0000020$	&	$1.515 \pm 0.008$	&	$1.6434 \pm 0.0052$	&	$62.47$	\\
\hline 
\sap{}		&	6	&	$0.9133795 \pm 0.0000037$	&	$1.022 \pm 0.010$	&	$1.8320 \pm 0.0097$	&	$44.69$	\\
\pdc{}		&	6	&	$0.9133824 \pm 0.0000032$	&	$1.023 \pm 0.009$	&	$1.8296 \pm 0.0085$	&	$44.99$	\\
\pap{}		&	6	&	$0.9133808 \pm 0.0000025$	&	$1.015 \pm 0.007$	&	$1.8362 \pm 0.0067$	&	$47.40$	\\
\hline 
\sap{}		&	7	&	$0.9318204 \pm 0.0000055$	&	$0.661 \pm 0.010$	&	$-0.6734 \pm 0.0145$	&	$31.42$	\\
\pdc{}		&	7	&	$0.9318298 \pm 0.0000047$	&	$0.662 \pm 0.008$	&	$-0.6730 \pm 0.0125$	&	$31.64$	\\
\pap{}		&	7	&	$0.9318241 \pm 0.0000035$	&	$0.669 \pm 0.006$	&	$-0.6675 \pm 0.0094$	&	$34.01$	\\
\hline 
\sap{}		&	8	&	$0.9678508 \pm 0.0000061$	&	$0.582 \pm 0.009$	&	$-1.3675 \pm 0.0160$	&	$29.83$	\\
\pdc{}		&	8	&	$0.9678491 \pm 0.0000052$	&	$0.585 \pm 0.008$	&	$-1.3487 \pm 0.0137$	&	$30.08$	\\
\pap{}		&	8	&	$0.9678499 \pm 0.0000038$	&	$0.586 \pm 0.006$	&	$-1.3552 \pm 0.0099$	&	$32.17$	\\
\hline 
\sap{}		&	9	&	$0.7942562 \pm 0.0000067$	&	$0.512 \pm 0.009$	&	$0.5684 \pm 0.0178$	&	$27.61$	\\
\pdc{}		&	9	&	$0.7942562 \pm 0.0000056$	&	$0.516 \pm 0.008$	&	$0.5502 \pm 0.0149$	&	$27.93$	\\
\pap{}		&	9	&	$0.7942592 \pm 0.0000041$	&	$0.509 \pm 0.005$	&	$0.5706 \pm 0.0108$	&	$29.53$	\\
\hline
\hline
\end{tabular}
\end{center}
\label{table: single-freq}
\end{table*}
	
	We continue extracting significant frequencies using iterative pre-whitening for each of the three light curves, and treat each extracted and optimised cosinusoid as independent of those that came before it, in order of decreasing amplitude (i.e. we perform a single-cosinusoid non-linear least squares fit at each iterative pre-whitening stage). We use amplitude S/N~$\geq 4.6$ as our significance criterion. The results of this single-cosinusoid non-linear least squares optimisation are shown in Table~\ref{table: single-freq}, for which only the first nine iterations are shown for clarity. We emphasise that the non-linear least-squares fit of an individual cosinusoid at a single pre-whitening iteration uses the residual light curve at that specific iteration (i.e. previous frequencies with larger amplitudes have already been subtracted from the light curve). The residual light curves and amplitude spectra after extracting the nine highest-amplitude frequencies are shown in Fig.~\ref{figure: iteration 9}. The residual light curves after so few iterations already show significant differences, despite the optimised cosinusoid parameters all being within $1\sigma$ of one another. 
	
\begin{figure}
\centering
\includegraphics[width=0.99\columnwidth]{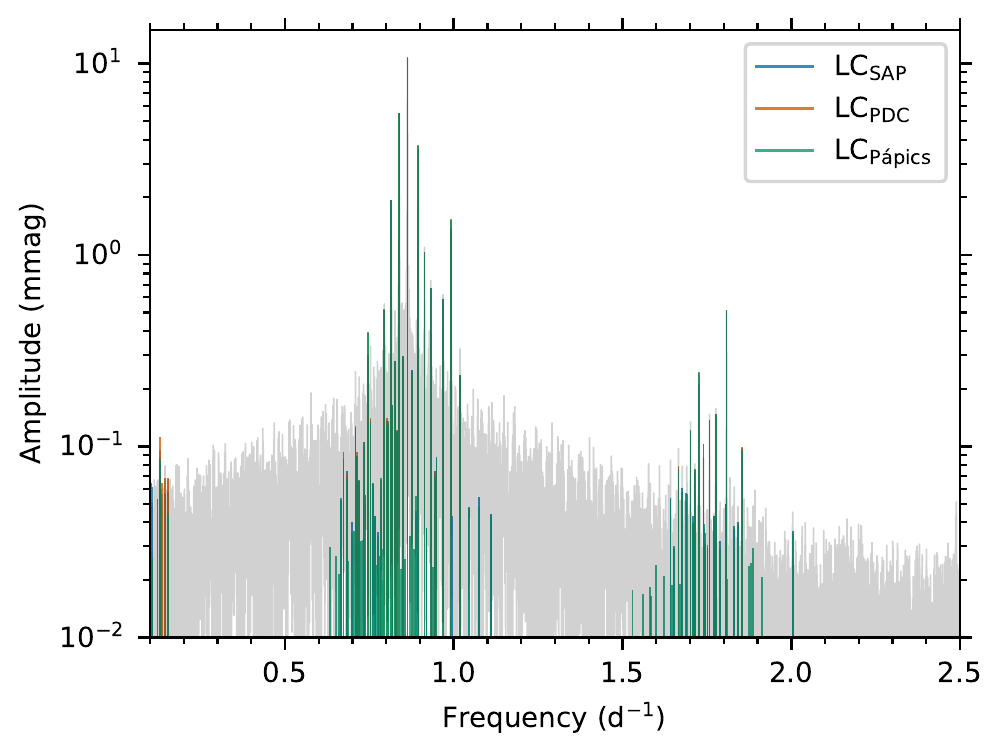}
\caption{Amplitude spectra of the optimised frequencies for KIC~7760680 using the single-cosinusoid solution for each of the \sap{}, \pdc{} and \pap{} light curves. The amplitude spectrum (grey) is calculated from the \pap{} light curve, such that low-amplitude modes only become significant above the noise after high-amplitude modes have been extracted during pre-whitening. Frequencies extracted below $\nu \leq 0.1$~d$^{-1}$ are dominated by instrumental systematics.}
\label{figure: extraction}
\end{figure}

	We show all the optimised frequencies within the low-frequency regime at the end of iterative pre-whitening in Fig.~\ref{figure: extraction}. Although there are some significant frequencies above 2.5~d$^{-1}$, the vast majority of significant frequencies lie within the $0.6 \leq \nu \leq 1.1$ and $1.6 \leq \nu \leq 2.0$~d$^{-1}$. As can be seen in Fig.~\ref{figure: extraction}, the significant frequencies in high-amplitude g-mode pulsators commonly appear in frequency groups, which can be explained by combination frequencies \citep{Kurtz2015b}. In this setup of treating each cosinusoid as independent of those that came before it, the respective frequencies and phases extracted from the three different light curves in the first $\sim$~50 iterations are consistent within $2\sigma$. At this stage of the iterative pre-whitening the extracted frequencies have $6 < {\rm S/N} < 10$. However, after about 50 iterations the frequency discrepancies are larger, and in some cases are as large as $7\sigma$. Furthermore, the extracted lists of frequencies from the three light curves are not the same, because the amplitude S/N~$\geq4.6$ criterion yields a different number (and extracted order) of frequencies for each light curve because of its underlying noise profile in the Fourier domain. 
	
	\subsubsection{Multi-cosinusoid non-linear least-squares fitting}
	\label{subsubsection: NL-LS}
		
	As a comparison, we also perform another iterative pre-whitening setup and extract significant frequencies using the \pap{} light curve of KIC~7760680 and a multi-cosinusoid non-linear least-squares fit with $N$ cosinusoids (cf. Eqn.~\ref{equation: cosine}) at each stage of the pre-whitening. In this second setup, we also use amplitude S/N~$\geq 4.6$ as our significance criterion. The results from this multi-cosinusoid non-linear least-squares fit setup are shown in Table~\ref{table: multi-freq}, in which the last column provides the frequency difference between the single- and multi-cosinusoid non-linear least-squares solutions, expressed in terms of the frequency uncertainty, $\sigma_{\nu}$, of the multi-cosinusoid solution. We note that some frequency labels are missing (e.g. 10 and 11), which is because they were identified as being within $2.5/\Delta\,T$ of a higher amplitude frequency hence were discarded and not included in the ultimate multi-cosinusoid fit owing to their potential spurious nature (see \citealt{Loumos1978}). Clearly, the resultant frequencies are discrepant in some cases as much as 15$\sigma$, but in others less than 1$\sigma$. The differences are not systematic with amplitude. The standard deviation of these frequency discrepancies is 4.2$\sigma$. 
	
	Although it is more computationally demanding, we conclude that such a numerical setup is important when extracting g-mode frequencies from space telescope light curves, since it includes the covariance of the frequencies, amplitudes and phases of all significant pulsation modes as opposed to only single pulsation mode being treated as independent of all others. Our conclusions are two-fold: 
	\begin{enumerate}
	\item the light curve extraction has a significant impact on the pulsation mode frequencies of a g-mode pulsator at the level of the spectral windows being sufficiently different enough to produce different pulsation mode frequencies, which gets systematically worse for lower amplitude pulsation modes; 
	\item the difference in extracted frequencies from a single-cosinusoid and multi-cosinusoid non-linear least-squares fit (cf. Eqn.~\ref{equation: cosine}) for a given light curve is representative of additional random uncertainty, which in some cases is much larger than the formal frequency precision. 
	\end{enumerate}
	The reason for the former conclusion can be understood as the analysis of a somewhat different data set yielding slightly different results. The reason for the latter conclusion is that the multi-cosinusoid non-linear least-squares fit includes the covariance matrix for all frequencies, amplitudes and phases in the determination of uncertainties.


	\subsection{Building period spacing patterns}
	\label{subsection: patterns}
	
	In this section we investigate the uncertainties in g-mode period spacing patterns considering two aspects: (i) the propagation of multi-cosinusoid non-linear least-squares uncertainties on the resultant period spacing pattern; and (ii) subjective selection from a list of extracted significant frequencies.
	
	\subsubsection{Statistical scatter in period spacing patterns}
	\label{subsubsection: statistical patterns}
	
	\begin{figure*}
	\centering
	\includegraphics[width=0.99\textwidth]{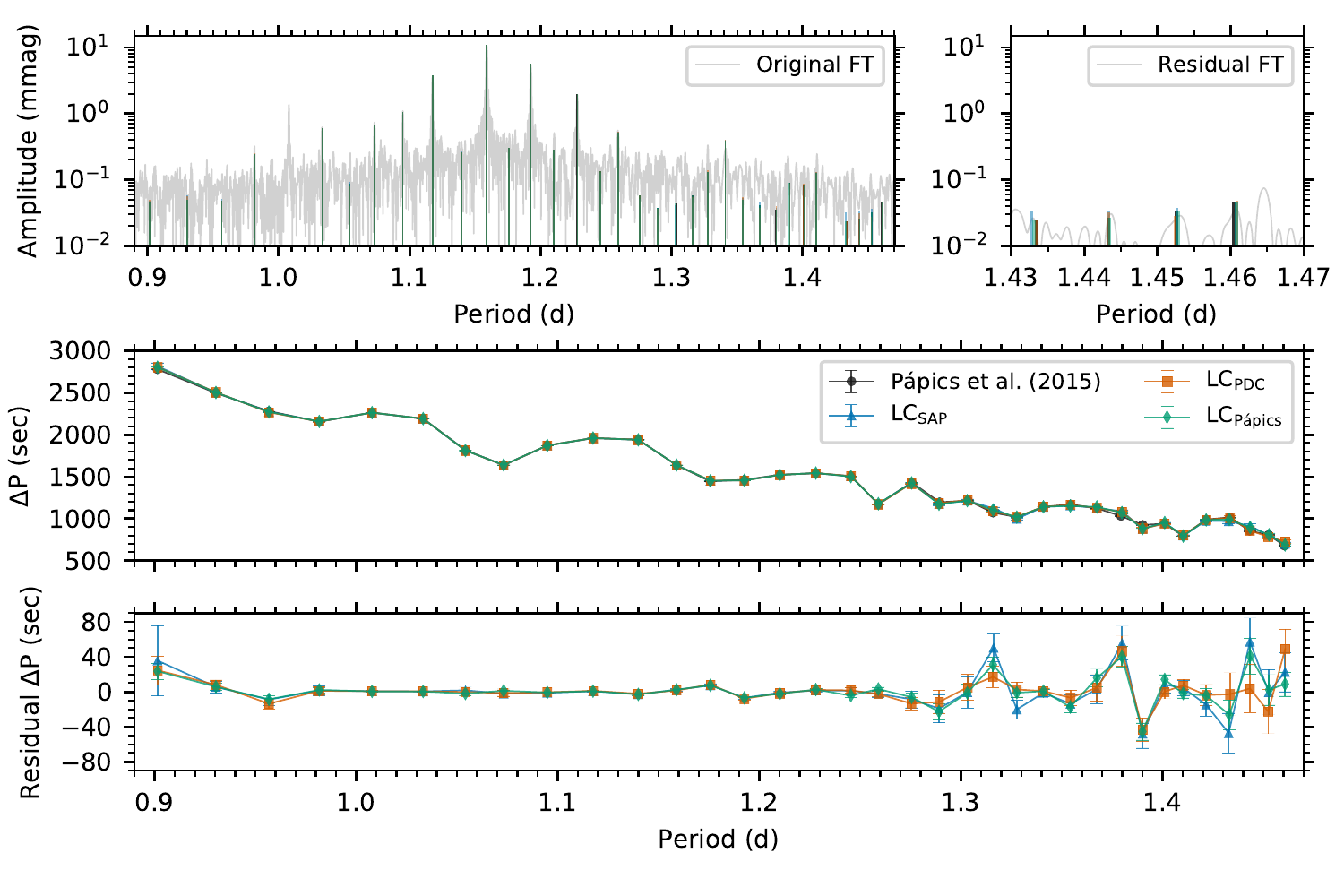}
	\caption{Comparison of different period spacing patterns for KIC~7760680, including the solution from \citet{Papics2015} based on the \pap{} light curve. The multi-cosinusoid non-linear least-squares fitted to the \sap{}, \pdc{} and \pap{} light curves are also shown on top of the amplitude spectrum in the top panels. The top-right panel is a zoom-in of the long-period regime. The resultant period spacing patterns of these four solutions are shown in the middle panel, with the residual of our patterns with that of \citet{Papics2015} shown in the bottom panel. Note that the formal uncertainties on the data in the middle panel are smaller than the symbol sizes.}
	\label{figure: papics pattern}
	\end{figure*}
	
	\begin{figure*}
	\centering
	\includegraphics[width=0.99\textwidth]{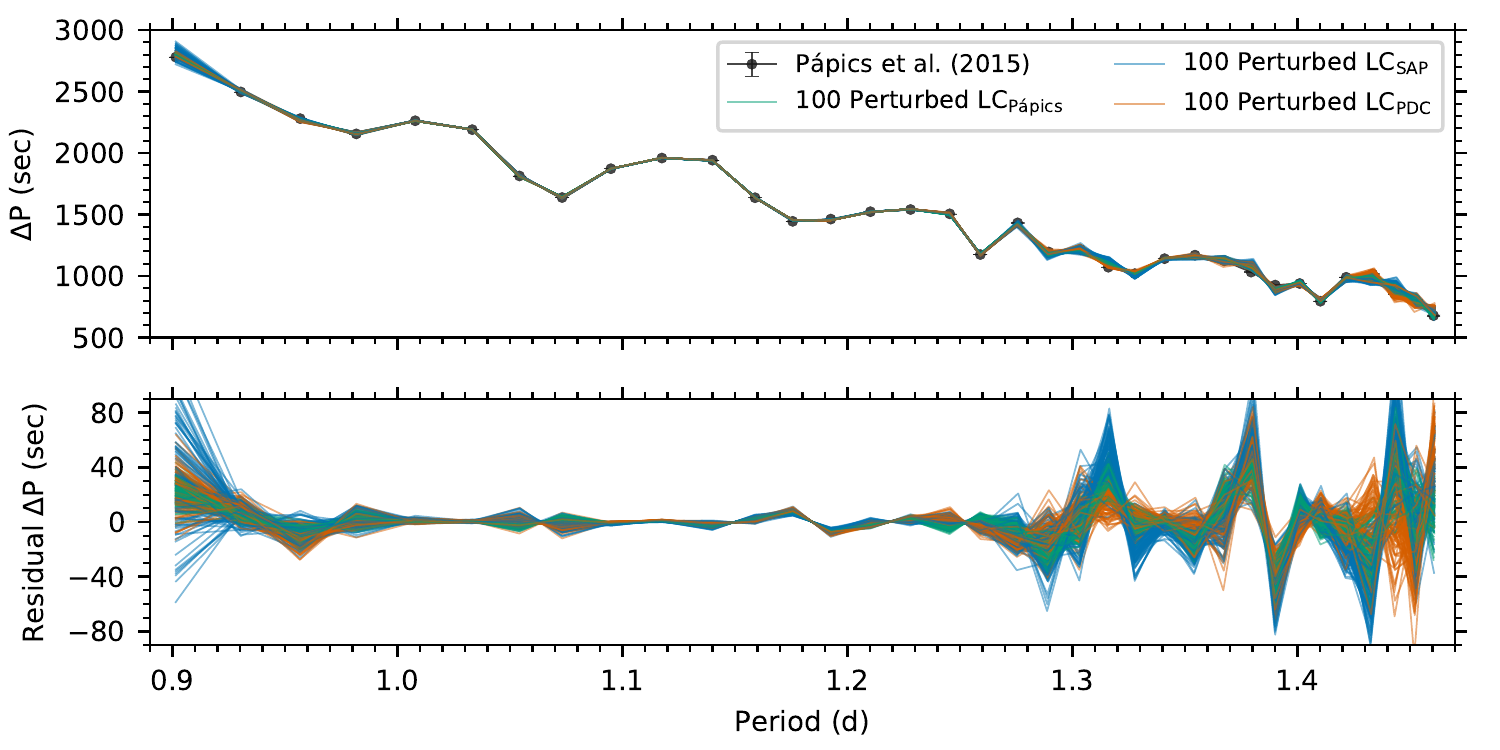}
	\caption{Resultant period spacing patterns arising from randomly perturbing pulsation mode frequencies within their 1$\sigma$ uncertainties 100 times based on the patterns shown in Fig.~\ref{figure: papics pattern} as input. The colour scheme is the same as in Fig.~\ref{figure: papics pattern}.}
	\label{figure: perturbed patterns}
	\end{figure*}
		
	To demonstrate the impact of a multi-cosinusoid non-linear least-squares fit, we compare the period spacing pattern extracted by \citet{Papics2015} assuming independent frequencies, with those resulting from our fits for the three \sap{}, \pdc{} and \pap{} light curves in Fig.~\ref{figure: papics pattern}. The differences among the three non-linear least-squares solutions is between $1-2\sigma$ in frequency. However, we find a much larger discrepancy between the solution of \citet{Papics2015} and that of our fits to the \pap{} light curve, which is between $0.1 \leq \sigma \leq 13$ and has a median value of $4.2\sigma$. This propagates into a median uncertainty of order 10~sec in the period spacing diagram, but the discrepancy is largest for the low amplitude modes (i.e. pulsation periods $> 1.28$~d). The main source of this uncertainty is that \citet{Papics2015} treated each frequency as independent during their iterative pre-whitening procedure, with their uncertainties determined using the idealistic formulae of \citet{Montgomery1999}, which assume parameters to be uncorrelated (see Appendix~A of \citealt{Breger1999f}). Therefore, the (lack of such a) fit represents an additional source of random uncertainty of, on average, 4.2$\sigma$ in frequency, and at least 10~sec in $\Delta$P values in the period spacing pattern, which is systematically larger for lower amplitude modes.
	
	As a second exercise, we generate three sets of 100 perturbed period spacing patterns by randomly perturbing the three \sap{}, \pdc{} and \pap{} patterns in Fig.~\ref{figure: papics pattern} within their $1\sigma$ frequency uncertainties. The resultant 100 period spacing patterns are shown in Fig.~\ref{figure: perturbed patterns}. We note that each set of 100 perturbed patterns are statistically significant and valid solutions, as they were all generated from the same input and $1\sigma$ uncertainties. Clearly, the propagation of the $1\sigma$ uncertainties in this Monte Carlo style sampling leads to a `cloud' of period spacing pattern solutions, whose scatter of up to 100~sec in the $\Delta$P ordinate axis of Fig.~\ref{figure: perturbed patterns} is a measure of the true uncertainty for KIC~7760680. We note that the scatter in Fig.~\ref{figure: perturbed patterns} is largest for the \sap{} light curve and smallest for the \pap{} light curve, which demonstrates the importance of optimising the light curve extraction methodology for g-mode asteroseismology.

	\subsubsection{Subjective component of building patterns}
	\label{subsubsection: subjective patterns}
	
	When manually building a g-mode period spacing pattern, one selects frequencies to maximise the number of consecutive radial orders, whilst adhering to some prior knowledge of what the average period spacing ought to be for a given star. This often leads to an inside-out approach to building patterns, such that the spacing between a few high-amplitude modes is replicated towards shorter and longer periods. This methodology was applied by \citet{Papics2015} when extracting the prograde dipole period-spacing pattern of KIC~7760680. This approach, however, is reliant on the expectation that period spacing patterns are smooth and periodic. It has recently been shown that resonances between overstable convective modes or inertial modes in the cores of massive stars with g~modes in the envelope can create sharp dips in period spacing patterns at different periods dependent on the rotation rate \citep{Ouazzani2020a, Saio2021a, Lee_U_2021a}. Also, the effect of a strong magnetic field within the convective cores of main sequence stars is predicted to perturb period spacing patterns in a mode-specific way \citep{Prat2019a, Prat2020a, VanBeeck2020a}. Therefore it is timely to revisit the extraction of g-mode period spacing patterns in KIC~7760680 allowing for such morphological effects.
	
	As can be seen in Figs~\ref{figure: papics pattern} and \ref{figure: perturbed patterns}, the high-amplitude modes between $0.98 \leq P \leq 1.23$~d have a relatively small uncertainty region, with their formal and Monte Carlo style sampling of the true uncertainty being $\sigma \left( \Delta{\rm P} \right) \simeq 10$~sec. For periods outside this range, much larger Monte Carlo uncertainties of order 100~sec are found. The inclusion of high-amplitude modes between $0.98 \leq P \leq 1.23$~d as members of the g-mode period spacing pattern is trivial; there are few alternatives. However, the choices from outside this range become more challenging. For example, Fig.~\ref{figure: papics pattern} shows that some neighbouring frequencies with higher amplitudes were not selected to be part of the pattern by \citet{Papics2015} for morphology reasons.
	
	\begin{figure*}
	\centering
	\includegraphics[width=0.99\textwidth]{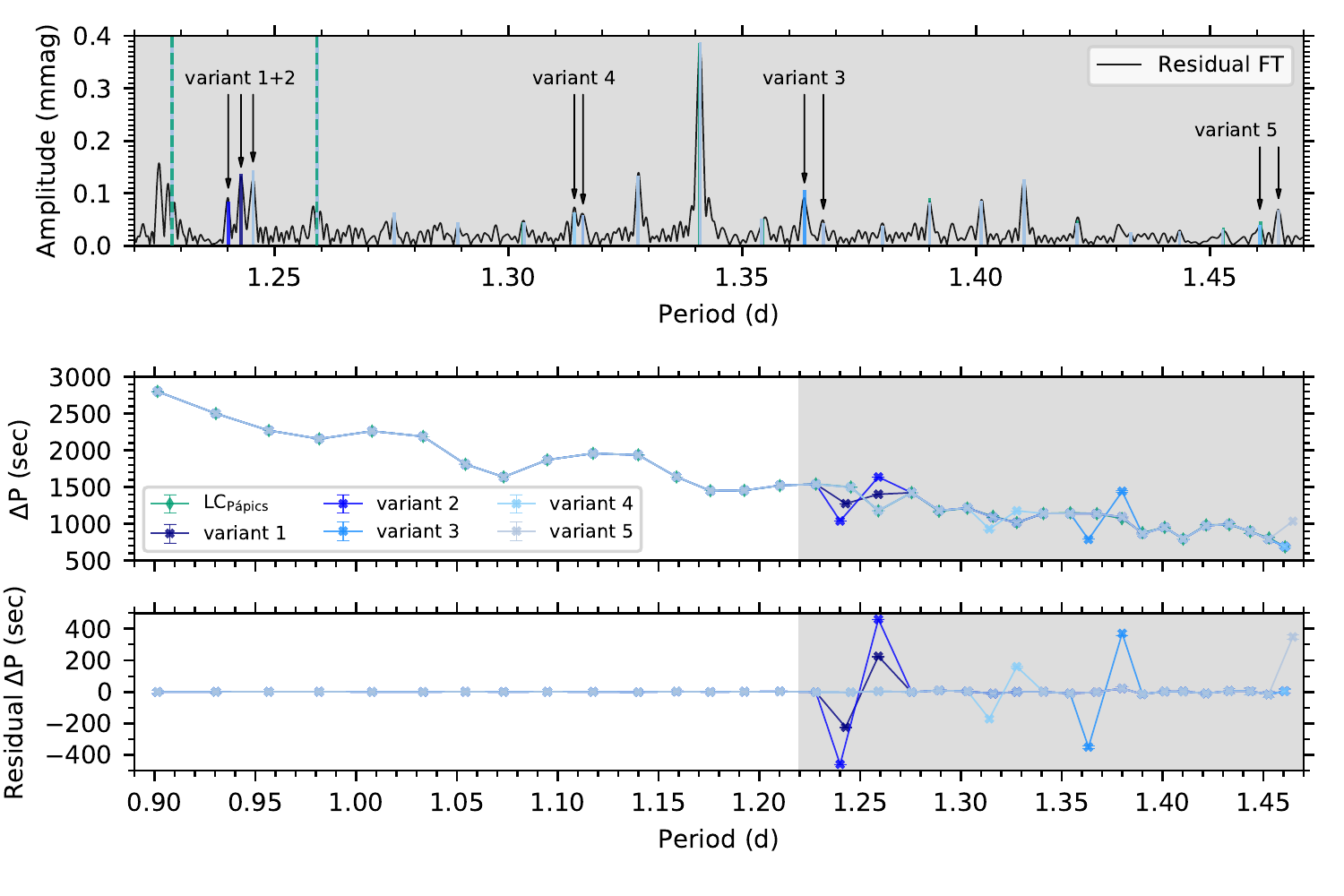}
	\caption{Top panel: zoom-in of the residual amplitude spectrum after the first nine pulsation mode frequencies have been removed to illustrate the choice of frequencies to include as part of a variant period spacing pattern. The location of two of the nine previously-extracted mode frequencies are within this period range and are denoted by dashed vertical green lines. Solid blue lines denote frequencies that have been chosen as part of one of the five variant period spacing patterns. Middle panel: the resultant period spacing patterns of our optimised \pap{} pattern is shown in green, and the additional five variant period spacing patterns are shown in blue. Bottom panel: residuals of the additional patterns with that of our optimised \pap{} benchmark pattern. The period regime of the top panel is denoted as the shaded grey region in the middle and bottom panels.}
	\label{figure: manual patterns}
	\end{figure*}
	
	To test the limitations of this inside-out approach when building period spacing patterns, we construct different patterns from the extracted list of significant frequencies using the multi-cosinusoid non-linear least-squares fit solution described in section~\ref{subsubsection: frequency comparison} and provided in Table~\ref{table: multi-freq}. We restrict ourselves to only the frequency list from the \pap{} light curve, because it is the highest quality. We also restrict ourselves to choosing different patterns from modes within in the low-frequency regime, because this is where the mode density is high and other significant frequencies are potential pattern members. This means that our choices of which significant frequencies to include and exclude within each different pattern are from the period range of $1.23 \leq P \leq 1.47$~d. In the case of KIC~7760680, the low-amplitude g~modes with periods above 1.27~d only become significant after several high-amplitude modes have been removed during iterative pre-whitening. In Fig.~\ref{figure: manual patterns}, we show a zoom-in of the residual amplitude spectrum after the first nine pulsation mode frequencies have been removed. 
	
	We build five `variant' period spacing patterns by selecting different significant frequencies to those of \citet{Papics2015} from the list of significant frequencies in Table~\ref{table: multi-freq} and show them in Fig.~\ref{figure: manual patterns}. We label five locations in Fig.~\ref{figure: manual patterns} at which a choice must be made on which frequency to include. For example, in variant patterns 1 and 2, we select the other frequencies from an apparent triplet of independent modes. In variant patterns 3 and 4, we select adjacent significant frequencies to those selected by \citet{Papics2015}. Finally, in variant pattern 5, we intentionally select an `incorrect' frequency at the end of the pattern to test the modelling sensitivity to the inclusion of such a low-amplitude long-period mode. For each of the five variant patterns, we also create a counterpart pattern for which the potentially ambiguous frequency is omitted entirely. Some of the five variant patterns may seem more appealing as solutions and others do not because they introduce glitches in the pattern morphology (e.g. variant pattern 3). This is because period spacing values are correlated with the period values so small differences in period can produce large differences in period spacings. The residuals between each of the five variant patterns and that of the optimised \pap{} solution is much larger than those in Fig.~\ref{figure: perturbed patterns} at locations where an alternative frequency has been introduced.


\section{Modelling Results}
\label{section: results}

Recently, \citet{Michielsen2021a} revisited the modelling of the SPB star KIC~7760680 using the MD as well as a classical $\chi^{2}$ merit function to extract properties including mass, age, metallicity, convective boundary and envelope mixing, using the period spacing pattern of \citet{Papics2015}. Forward asteroseismic modelling is at least a 7D minimisation problem, with the interior rotation rate being one of these parameters \citep{Aerts2018b}. Precise rotation rates of main-sequence stars can be measured from the tilt in g-mode period spacing patterns \citep{Aerts2021a}. Hence the determination of the rotation frequency a priori from the g-mode period spacing pattern reduces the dimensionality of the subsequent modelling. This motivated \citet{Michielsen2021a} to fix the rotation frequency of KIC~7760680 and assume a uniform interior rotation profile with a rotation frequency of 0.4805~d$^{-1}$, with this value being determined from the previous modelling of \citet{Moravveji2016b}. As our study is the observational counterpart to their theoretical study, we also assume the same rotation frequency as a fixed input in forward asteroseismic modelling and the exact same numerical setup as employed by \citet{Michielsen2021a} for comparison purposes.

We use the `radiative' grid setup of \citet{Michielsen2021a}, which includes stellar structure and evolution models calculated using {\sc MESA} (r12115; \citealt{Paxton2011, Paxton2013, Paxton2015, Paxton2018, Paxton2019}). We also make use of the nested-parameter setup of \citet{Michielsen2021a}, such that fitting can be done using 4, 5 or 6 free parameters whilst penalising for the increased number of free parameters. In the 4-parameter setup, the free parameters are stellar mass ($M$), metal mass fraction ($Z$), core hydrogen content ($X_c$) and envelope mixing ($\log\,D_{\rm env}$) using the expected profile for internal gravity waves (i.e. $D_{\rm env} \propto \rho^{-1}$) as predicted based on hydrodynamical simulations by \citet{Rogers2017c}. In the 5-parameter setup, the addition of core-boundary mixing (CBM) using a diffusive exponential prescription ($f_{\rm CBM}$) is included. In the 6-parameter setup, a second CBM parameter is included, specifically the extent of a step profile above the convective core ($\alpha_{\rm CBM}$). The reason for including $f_{\rm CBM}$ and $\alpha_{\rm CBM}$ as separate parameters is to include a larger range and diversity of core masses in the model grid, which g-modes directly probe \citep{Michielsen2019a, Michielsen2021a}. 

From prior testing, the {\sc MESA} model grid of \citet{Michielsen2021a} was computed to finely sample the possible parameter space of KIC~7760680 based on its g-mode period spacing pattern, far beyond its 3$\sigma$ spectroscopic error box in the HR~diagram. Mass values range between $2.8 \leq M \leq 3.7$~M$_{\odot}$ in steps of 0.1~M$_{\odot}$, core hydrogen content ranges between $0.30 \leq X_{\rm c} \leq 0.60$ in steps of 0.02 and three metallicity values of $Z =$~0.015, 0.019 and 0.023 are included, which covers the spectroscopic uncertainties, with the corresponding helium mass fraction calculated using an enrichment law for each $Z$ value and imposing $X+Y+Z=1$. Interior mixing parameters include: $0.000 \leq f_{\rm CBM} \leq 0.030$ in steps of 0.005, $0.00 \leq \alpha_{\rm CBM} \leq 0.30$ in steps of 0.05, and $0.0 \leq \log\left(D_{\rm env}\right) \leq 2.0$ in steps of 0.5. For each {\sc MESA} model, the corresponding adiabatic prograde dipole g-mode frequencies using the pulsation code {\sc GYRE} (v5.2; \citealt{Townsend2013b}) were calculated assuming the pre-determined rotation frequency of 0.4805~d$^{-1}$ and uniform interior rotation. We refer the reader to \citet{Michielsen2021a} for full details of the {\sc MESA} grid calculation and the {\sc GYRE} calculations. 

The method of forward asteroseismic modelling involves using a maximum likelihood estimator and a quantitative fit of observables to the theoretical counterparts from the grid of models, which in this case is the pulsation periods within a g-mode period spacing pattern. Recent forward asteroseismic modelling studies have moved beyond a simplistic $\chi^2$ merit function in favour of using the MD \citep{Aerts2018b}, which allows non-linear correlations between fitting parameters and observables to be included, as well as incorporating theoretical uncertainties in the form of the full variance-covariance matrix for the observables and the theoretical predictions in addition to observational uncertainties. In the following subsections, we employ the same methodology of \citet{Michielsen2021a} and perform quantitative fitting between observed pulsation periods to those predicted from a grid of stellar structure models using the MD as merit function. In each instance, for the purpose of making our results comparable to those of \citet{Michielsen2021a}, the best model within the grid is defined as that with the lowest MD value from within the 3$\sigma$ spectroscopic error box of KIC~7760680 (cf. Table~\ref{table: spectra}), and the same grid of models is used.


	\subsection{Modelling perturbed patterns}
	\label{subsection: results: perturbed patterns}
	
	In this section, we empirically test if the quality of the extracted light curve of KIC~7760680 and the scatter within the resultant period spacing pattern clouds are large enough to impact the resultant best fitting model parameters. The g-mode period spacing pattern of \citet{Papics2015}  is compared to the 100 perturbed patterns derived from the same light curve (i.e. \pap{}) from Fig.~\ref{figure: perturbed patterns}. Furthermore, the differences between the three sets of 100 perturbed patterns for the \sap{}, \pdc{} and \pap{} perturbed patterns provide an insight on the impact of the light curve quality and effect of a multi-cosinusoid non-linear least-squares fit on the model parameters.

	\subsubsection{Scatter in best-fitting models}
	
\begin{figure}
	\centering
	\includegraphics[width=0.99\columnwidth]{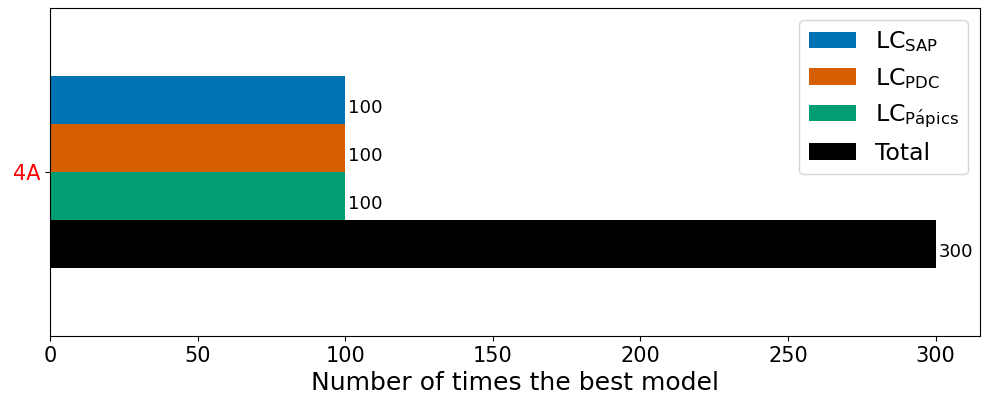}
	\includegraphics[width=0.99\columnwidth]{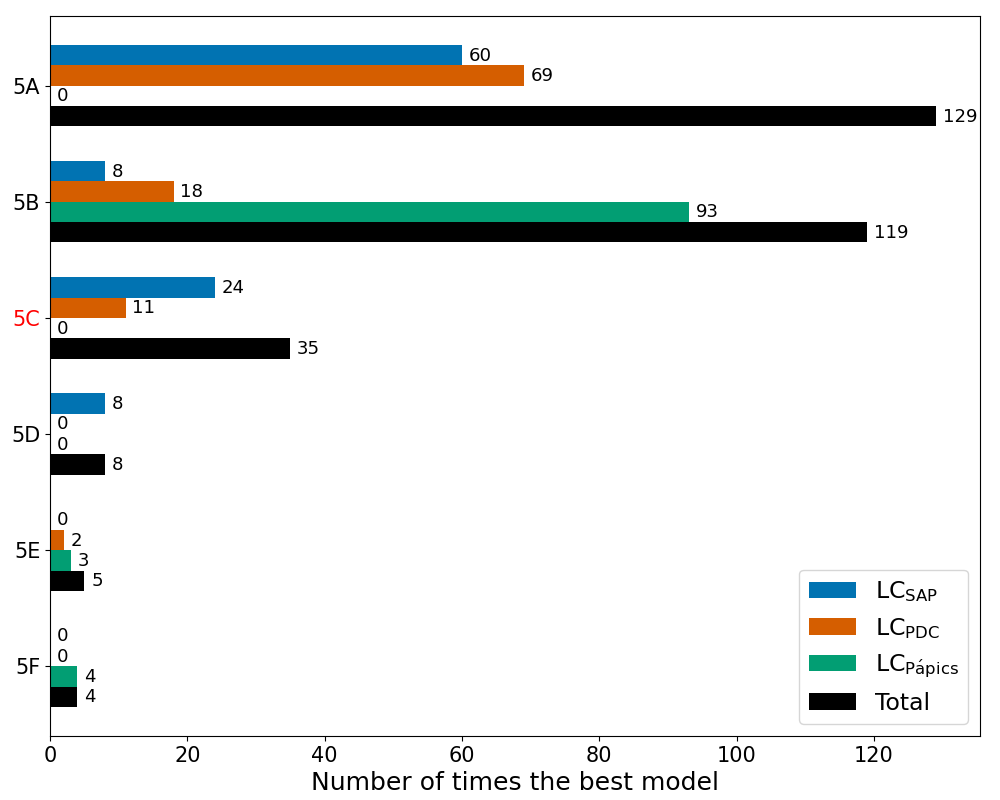}
	\includegraphics[width=0.99\columnwidth]{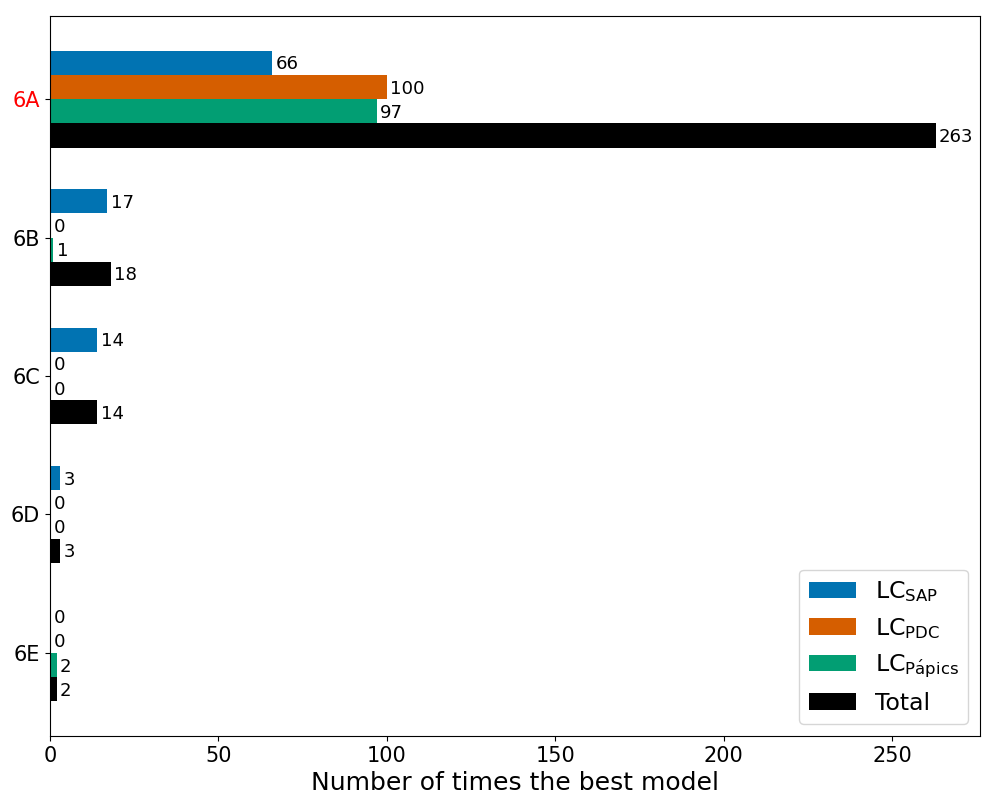}
	\caption{Histograms showing the number of times each best model was returned from the three sets of 100 perturbed g-mode period spacing patterns derived from the \sap{}, \pdc{} and \pap{} light curves. The parameters of the models are given in Table~\ref{table: best models}. Solutions based on a 4-, 5- and 6-parameter setup are shown in the top, middle and bottom panels, respectively. The model names in red (i.e. 4A, 5C and 6A) are the best models determined by \citet{Michielsen2021a} using the g-mode period spacing pattern from \citet{Papics2015}.}
	\label{figure: perturbed histograms}
	\end{figure}

	The first question is whether the quality of the light curves and the resultant g-mode period spacing pattern(s) have large enough uncertainties to produce a significantly different set of best model parameters. We employ the 4-, 5- and 6-parameter nested model selection methodology of \citet{Michielsen2021a} for each pattern from the three sets of 100 perturbed patterns for KIC~7760680 and determine the best model from within the grid in each case. Our results are summarised as histograms in Fig.~\ref{figure: perturbed histograms} for each setup, with the parameters for each of the models specified in Table~\ref{table: best models}. All 300 perturbed patterns return the same best model denoted as `4A' in the 4-parameter setup. Interestingly, this is the same best model as found by \citet{Michielsen2021a} when using the pattern of \citet{Papics2015} and the 4-parameter setup. Hence when only considering a 4-parameter setup (i.e. $M$, $Z$, $X_c$ and $\log\,D_{\rm env}$) the impact of the quality of extracted light curve and the observational uncertainties can be considered small compared to the theoretical uncertainties in the case of KIC~7760680, such that they do not significantly influence the selection of the best model in our MD framework.
	
\begin{table}
\caption{Parameters of the best models for the 4-, 5- and 6-parameter setups given in Figs~\ref{figure: perturbed histograms} and \ref{figure: best models}.} 
\begin{center}
\begin{tabular}{c c c c c c c}
\hline \hline
&	$M$	&	$Z$	&	$\log(D_{\rm env})$	&	$\alpha_{\rm CBM}$	&	$f_{\rm CBM}$	&	$X_{\rm C}$	\\
&	(M$_{\odot}$)	&	$$	&	$$	&	$$	&	$$	&	$$	\\
\hline
4A	&	3.4	&	0.019	&	1.5	&	--	&	--		&	0.44	\\
\hline
5A	&	3.2	&	0.015	&	0.0	&	--	&	0.020	&	0.46	\\
5B	&	3.4	&	0.023	&	0.0	&	--	&	0.005	&	0.40	\\
5C	&	3.4	&	0.023	&	1.0	&	--	&	0.020	&	0.46	\\
5D	&	3.5	&	0.023	&	0.5	&	--	&	0.020	&	0.46	\\
5E	&	3.4	&	0.019	&	0.5	&	--	&	0.010	&	0.42	\\
5F	&	3.3	&	0.023	&	0.5	&	--	&	0.005	&	0.40	\\
\hline
6A	&	3.5	&	0.023	&	0.0	&	0.20	&	0.000	&	0.44	\\
6B	&	3.4	&	0.023	&	0.5	&	0.05	&	0.000	&	0.40	\\
6C	&	3.4	&	0.019	&	0.5	&	0.20	&	0.000	&	0.46	\\
6D	&	3.0	&	0.019	&	0.0	&	0.25	&	0.010	&	0.46	\\
6E	&	3.3	&	0.023	&	0.0	&	0.05	&	0.005	&	0.40	\\
\hline \hline
\end{tabular}
\end{center}
\label{table: best models}
\end{table}

	\begin{figure}
	\centering
	\includegraphics[width=0.99\columnwidth]{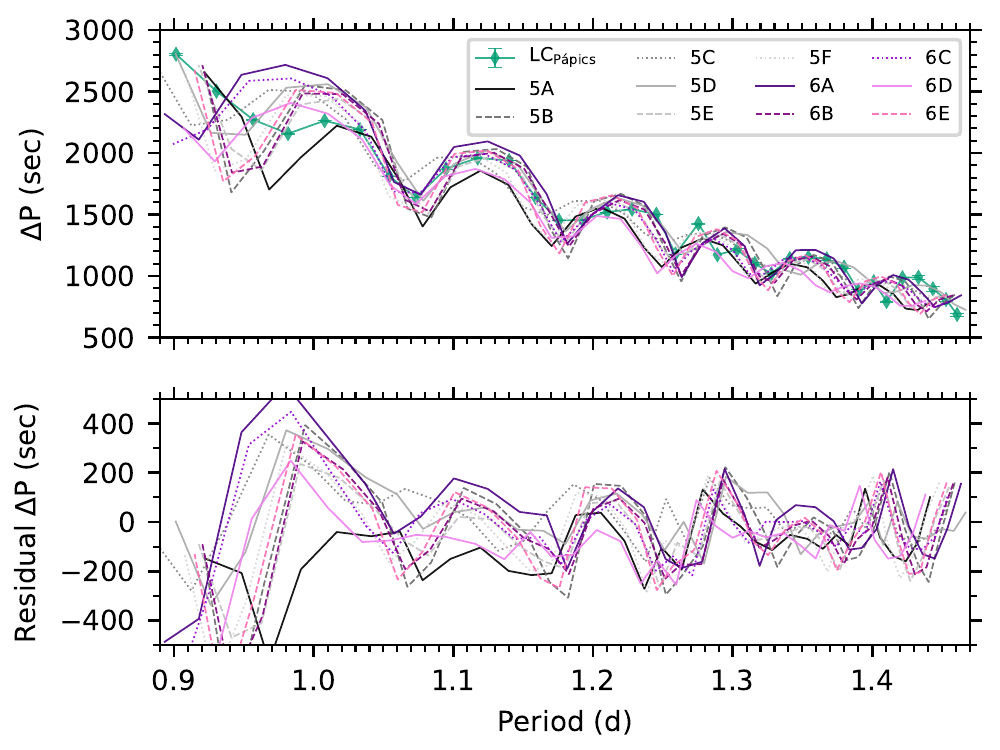}
	\caption{Top panel: best fitting period spacing patterns for best models returned using the 5- and 6-parameter setup in Fig.~\ref{figure: perturbed histograms}. The parameters of the models are given in Table~\ref{table: best models}. The optimised pattern based on the \pap{} light curve (cf. Fig.~\ref{figure: papics pattern}) is included to guide the eye. Bottom panel: residuals between each best-fitting theoretical pattern and the optimised pattern based on the \pap{} light curve.}
	\label{figure: best models}
	\end{figure}
	
	However, this is not the case when using the 5-parameter setup (i.e. addition of $f_{\rm CBM}$ as a fitting parameter), as can be seen in Fig.~\ref{figure: perturbed histograms}. Since g~modes are particularly sensitive to the size of the convective core, the addition of $f_{\rm CBM}$ as a free parameter allows for a more diverse range of theoretical patterns to fit to the observations. For example, larger CBM in the form of $f_{\rm CBM}$ directly modifies the size of the `dips' in a period spacing pattern \citep{Miglio2008a}, hence allowing for more flexibility in fitting the morphology of a period spacing pattern. Our modelling results in the 5-parameter setup, summarised in Fig.~\ref{figure: perturbed histograms}, show that most of the perturbed patterns prefer a different best model to that of \citet{Michielsen2021a}. The difference in the best-fitting parameters may seem small at first glance, but it is important to remember that the MD is maximum-likelihood point estimator and the uncertainty region for parameters of a given solution depend on the best model and the underlying distribution in MD values for each fit \citep{Michielsen2021a, Mombarg2021a}. Therefore, these results demonstrate two conclusions when using the common 5-parameter setup in forward asteroseismic modelling: (i) the use of a multi-cosinusoid non-linear least-squares fit in determining which frequencies are extracted and included in g-mode period spacing pattern impacts the resultant best model parameters; and (ii) the quality of the light curve impacts the morphology of the period spacing pattern (both in the periods and their uncertainties) such that different best-fitting model parameters are found.
	
	It is typical to use the 5-parameter setup (i.e. including only a single parameter, such as $f_{\rm CBM}$ or $\alpha_{\rm CBM}$ to control the amount of CBM) in modelling studies of stars with g-mode period spacing patterns \citep{Moravveji2015b, Moravveji2016b, Szewczuk2018a, Mombarg2019a, Pedersen2021a}. However, \citet{Michielsen2021a} went beyond such a setup and allowed for an additional parameter, such that the combination of both $f_{\rm CBM}$ and $\alpha_{\rm CBM}$ controlled the total amount and shape of CBM. Interestingly, in this 6-parameter setup, the vast majority of best fitting models from our three sets of 100 perturbed patterns return the exact same model as found by \citet{Michielsen2021a} using the g-mode period spacing pattern of \citet{Papics2015}. This demonstrates that the addition of a second parameter in modelling CBM in main sequence g-mode pulsators is more robust against unaccounted for uncertainties in the observations. In other words, in a 5-parameter modelling setup, the theoretical model grid does not contain the same flexibility to take into account the shape of the observed period spacing pattern and its uncertainties, such that observed periods perturbed within their uncertainties return different results, which are of order 10\% for mass. However, in the 6-parameter setup, the theoretical model grid is more robust and provides more consistent fits that can better handle potential discrepancies that may or may not have been introduced during the light curve reduction and iterative pre-whitening.
	
	The theoretical period spacing patterns for the best fitting models are shown in Fig.~\ref{figure: best models}. It is clear that the variance in the residual theoretical patterns with the optimised observed pattern in Fig.~\ref{figure: best models} is more than 200~sec, which is larger than the $\sim50$~sec variance among the perturbed observed patterns in Fig.~\ref{figure: perturbed patterns}. Therefore, the theoretical uncertainties associated with different input physics in stellar models are dominant in forward modelling. Yet, the observational uncertainties are not negligible and contribute to the scatter in Fig.~\ref{figure: best models}, specifically because the perturbed locations and depths of dips in the observed patterns are sufficient to sometimes return a different best model. We note that largest theoretical uncertainties are associated with short-period low-radial order g~modes, whereas the largest observational uncertainties are associated with the low-amplitude large-radial order g~modes.

	\subsubsection{Impact on inferred core masses}
		
\begin{figure}
	\centering
	\includegraphics[width=0.99\columnwidth]{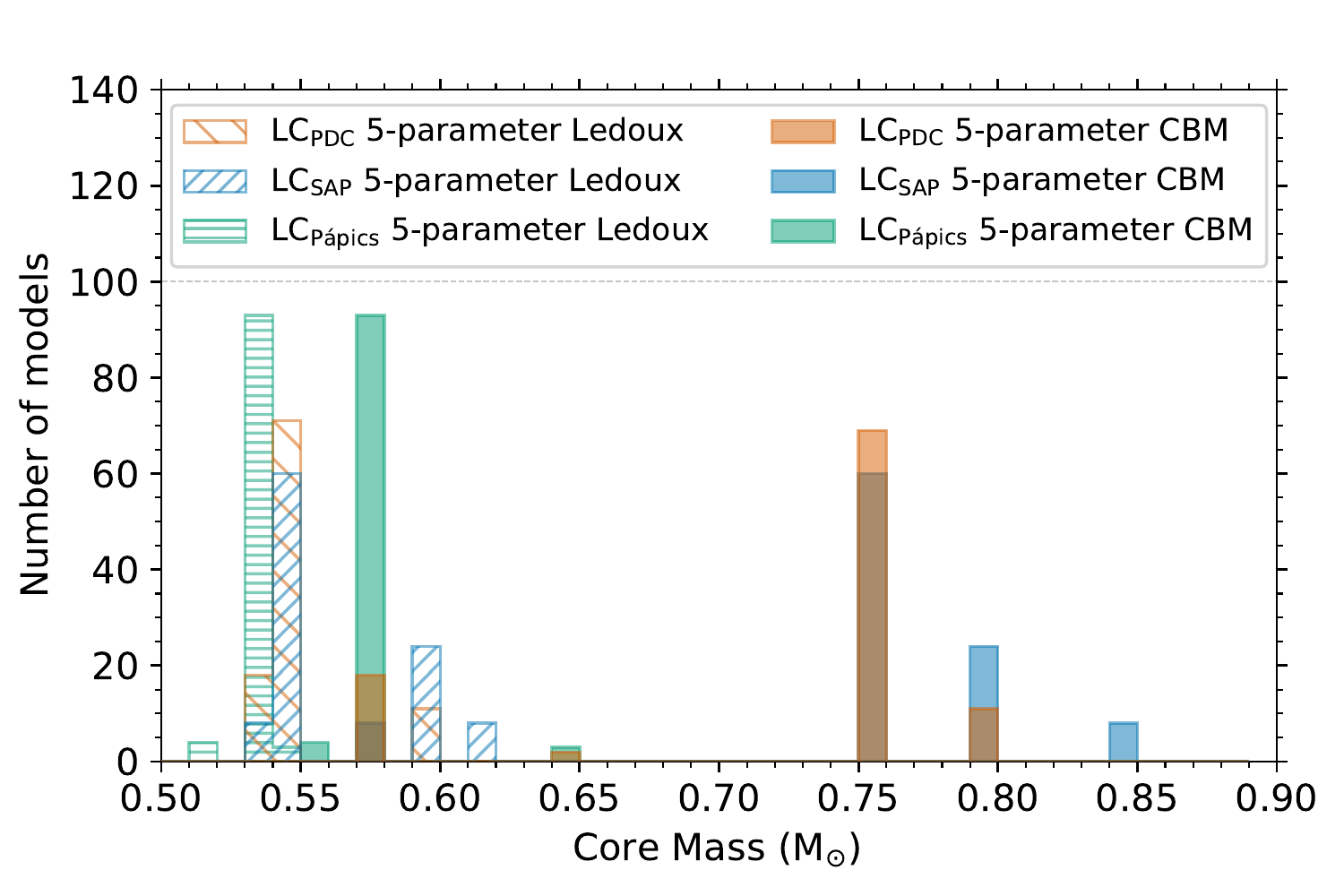}
	\includegraphics[width=0.99\columnwidth]{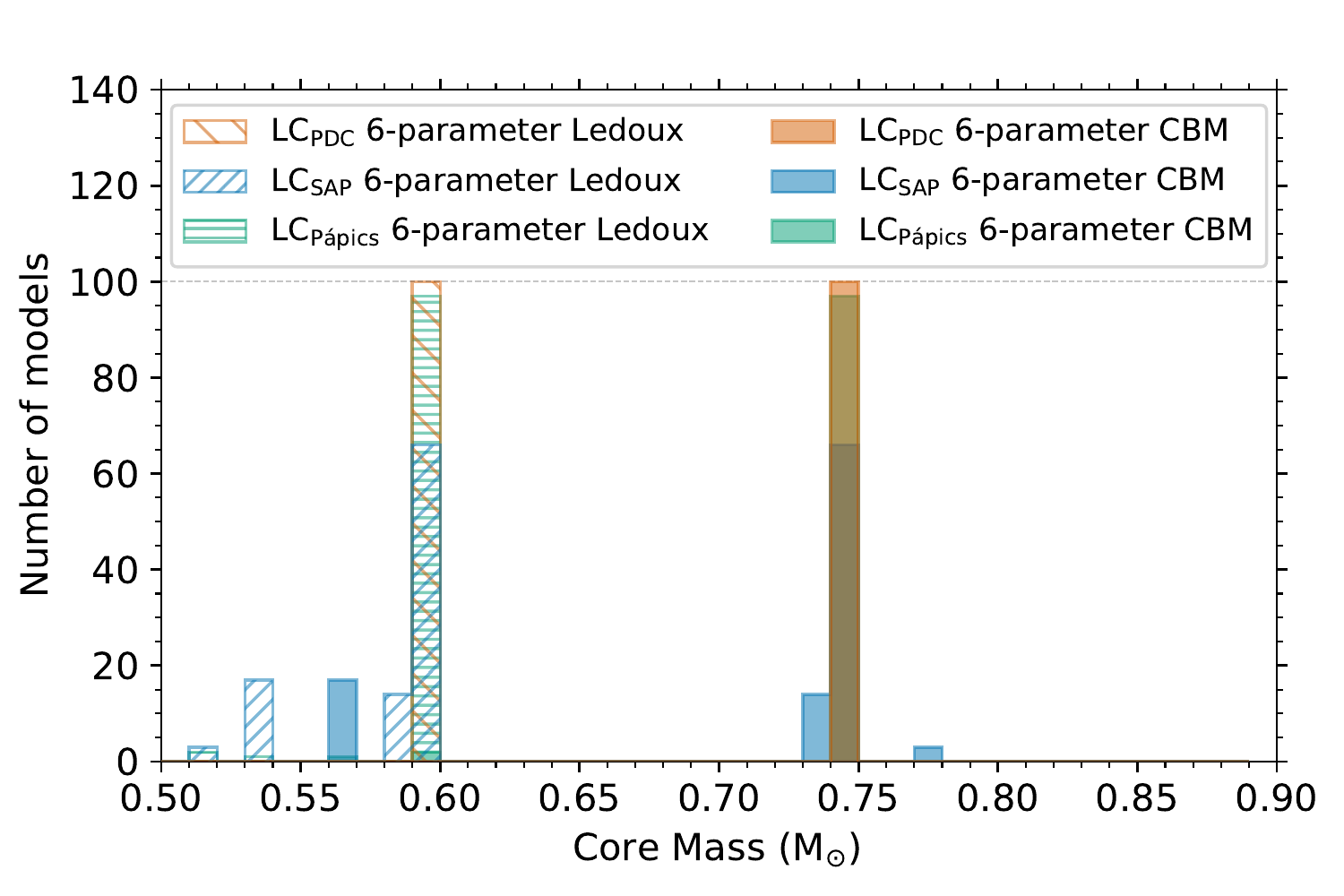}
	\caption{Histograms for the inferred core masses of the three sets of 100 perturbed period spacing patterns for the \sap{}, \pdc{} and \pap{} light curves. The top and bottom panels show the distribution for the 5- and 6-parameter setups. Two forms of the inferred core mass are provided, one defined as the mass within the region that satisfies the Ledoux criterion, and the second which additionally includes the mass within the core-boundary mixing (CBM) region as well.}
	\label{figure: core mass histograms}
	\end{figure}

	The second question is whether the observed scatter in the best fitting models and resultant theoretical period spacing patterns (cf. Fig.~\ref{figure: best models}) for the 5- and 6-parameter solutions produces a significant difference in the inferred core mass compared to the uncertainty region for an individual model. For the both the 5- and 6-parameter setups, we determine the core mass for each of the three sets of 100 perturbed period spacing patterns from the \sap{}, \pdc{} and \pap{} light curves and show them in Fig.~\ref{figure: core mass histograms}. We use two formulations for the inferred core mass: (i) the mass contained within the region that satisfies the Ledoux criterion; and (ii) that of (i) and also the mass within the overlying CBM region. 
	
	In Fig.~\ref{figure: core mass histograms} we demonstrate the larger scatter in the inferred core mass values (both with and without the inclusion of the CBM region) in the 5-parameter setup compared to the 6-parameter setup for the three sets of 100 perturbed patterns. For example, in the 6-parameter setup, more than 65\% of the 100 \sap{} and almost all of the \pdc{} and \pap{} patterns have the same inferred core mass when including or excluding the CBM region. On the other hand, in the 5-parameter setup, there is less agreement. The inclusion of the CBM region when determining the core mass has a much larger effect when using the 5-parameter setup. For example, the majority of the \pap{} patterns prefer a core mass of approximately 0.58~M$_{\odot}$ including the CBM region, but the \sap{} and \pdc{} patterns prefer core masses in excess of 0.75~M$_{\odot}$. We note that the masses among the best models range between $3.0 \leq M \leq 3.5$~M$_{\odot}$, such that the scatter in the fractional core masses in the 5-parameter setup are also much larger compared to the 6-parameter setup.
	
	To judge whether the width of the distribution in core mass is significant, we compare it to the uncertainty for the inferred core mass from an individual model. For simplicity, we focus on the modelling results based on the 100 perturbed patterns from only the \pap{} light curve, because they are superior to the \sap{} and \pdc{} solutions. From the 100 perturbed patterns based on the \pap{} light curve, we find that the 2$\sigma$ core mass uncertainties range between 0.01 and 0.1~M$_{\odot}$. This corresponds to a fractional uncertainty up to 20\% dependent on the model. On the other hand, when looking at the 6-parameter setup, the distribution in inferred core masses is much narrower and the uncertainties are much smaller as well. In most of the 100 perturbed patterns based on the \pap{} light curve, the uncertainty on the core mass in the 6-parameter is much smaller than the grid step size, such that only the best model is returned within its own 2$\sigma$ confidence interval. Hence the fractional uncertainties on core mass in the 6-parameter setup are of order 1\% compared to the 20\% in the 5-parameter setup. This further demonstrates the robustness of the 6-parameter setup and the MD modelling framework against potentially unaccounted for observational uncertainties.
	
	\subsubsection{Parameter confidence intervals}
	
\begin{figure*}
	\centering
	\includegraphics[width=0.99\textwidth]{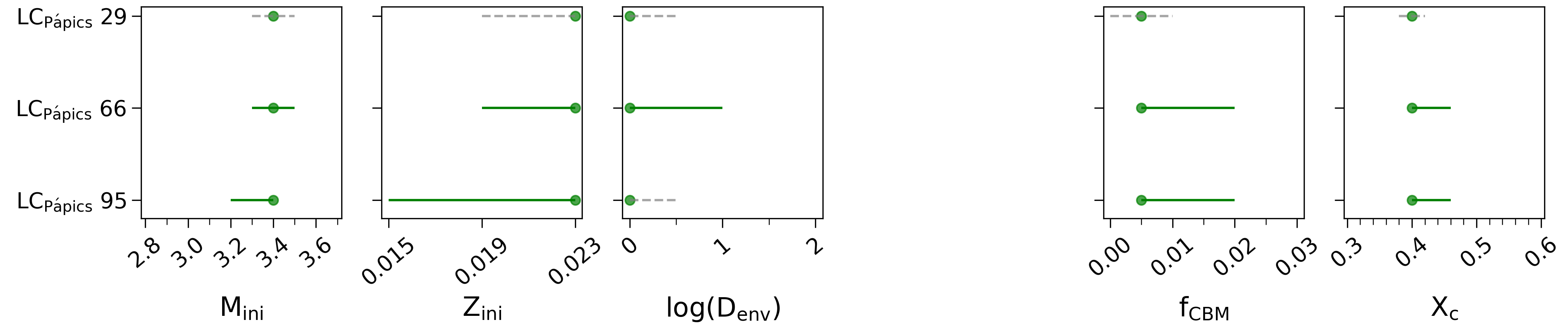}
	\includegraphics[width=0.99\textwidth]{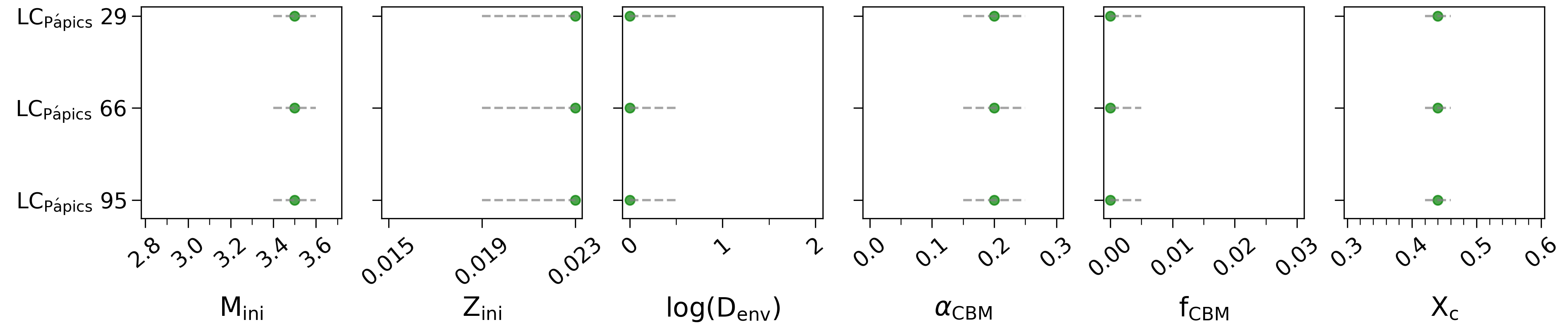}
	\caption{Parameters and 2$\sigma$ confidence intervals for the three best fitting \pap{} perturbed period spacing patterns shown as filled circles and solid lines, respectively. If a confidence interval calculation using Bayes’ theorem did not return any additional models, an upper limit estimate is taken to be the grid step size for a given parameter, which is indicated by the dashed grey lines.}
	\label{figure: confidence intervals}
	\end{figure*}

	Following \citet{Michielsen2021a}, we also calculate confidence intervals for model parameters in each setup using Bayes’ theorem and the distribution of likelihoods returned from the forward asteroseismic modelling results. Figure~\ref{figure: confidence intervals} shows the value of each parameter and its 2$\sigma$ confidence interval for the 5- and 6-parameters for the three best fitting (i.e. lowest MD value) perturbed period spacing patterns from the set of 100 for the \pap{} light curve. Similarly to \citet{Michielsen2021a}, if the application of Bayes’ theorem returns no other models within the 2$\sigma$ confidence interval based on the model's likelihood, the uncertainty is taken to be smaller than the grid step size for a given parameter. 
	
	For the 5-parameter setup, Fig.~\ref{figure: confidence intervals} shows that the confidence interval for each individual parameter is quite different from one pattern to another, but generally consistent within 2$\sigma$, but the inferred core masses generally are not (cf. Fig.~\ref{figure: core mass histograms}). Whereas for the 6-parameter setup, the best three models all return the same model parameter values, and have parameter uncertainties smaller than the step size of the grid. Our results indicate that asteroseismic modelling studies of g-mode period spacing patterns are most robust against unaccounted for observational uncertainties when using the 6-parameter setup in a MD framework.
			
	\subsection{Modelling subjective patterns}
	\label{subsection: results: subjective patterns}
	
\begin{figure}
	\centering
	\includegraphics[width=0.99\columnwidth]{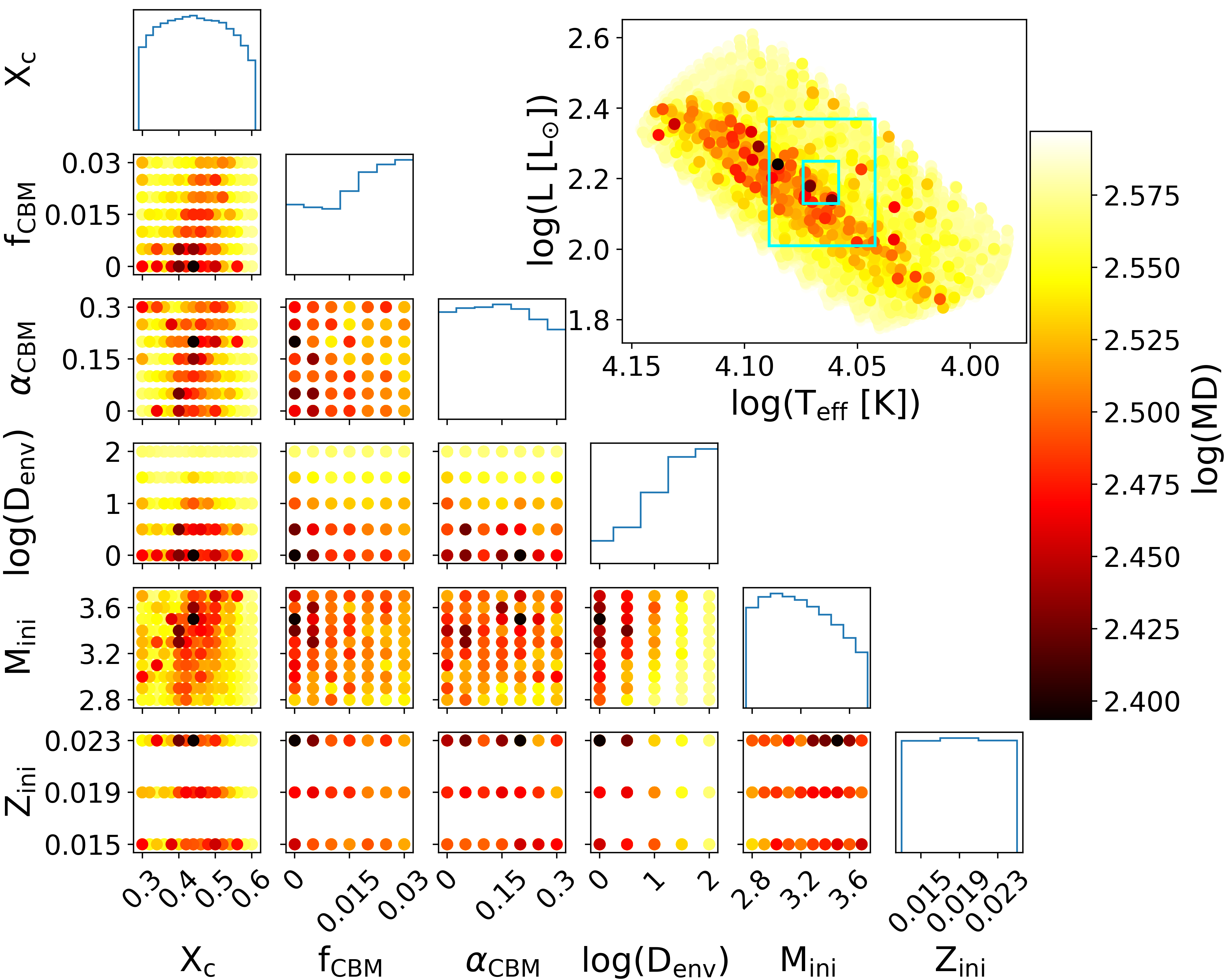}
	\caption{Summary figure for the modelling results of the benchmark period spacing pattern, which is the \citet{Papics2015} pattern optimised using a multi-cosinusoid non-linear least-squares fitted to the \pap{} light curve (cf. Fig.~\ref{figure: papics pattern}). The cyan boxes in the HR~diagram denote the 1- and 3-$\sigma$ spectroscopic error box for KIC~7760680.}
	\label{figure: benchmark modelling}
	\end{figure}

	In this subsection, we investigate the impact of manually selecting (ambiguous) frequencies as part of a period spacing pattern on the modelling results. We define a benchmark period spacing pattern for comparison purposes. Since it was demonstrated that the period spacing pattern was far more robust and less prone to random frequency uncertainties caused by choices in the light curve extraction and pre-whitening, we choose the pattern optimised using a multi-cosinusoid non-linear least-squares fitted to the \pap{} light curve, which is shown in Figs~\ref{figure: papics pattern} and \ref{figure: manual patterns}. As in the previous section, the rotation frequency is kept fixed for all of the subsequent modelling setups. Since the changes in individual frequencies among the variant patterns are extremely small, we checked that the same rotation frequency was derived from the tilt of the g-mode period spacing pattern for all patterns within its uncertainties. A summary of the modelling results for the benchmark solution is shown in Fig.~\ref{figure: benchmark modelling}, in which the resultant 2D and 1D parameter probability distributions are provided, as well as the distribution of all models in the HD~diagram colour-coded by their MD value. We note that the distribution of best models may appear sporadic, but this is a result of theoretical uncertainties contained within the modelling grid \citep{Michielsen2021a}.

\begin{figure*}
	\centering
	\includegraphics[width=0.99\textwidth]{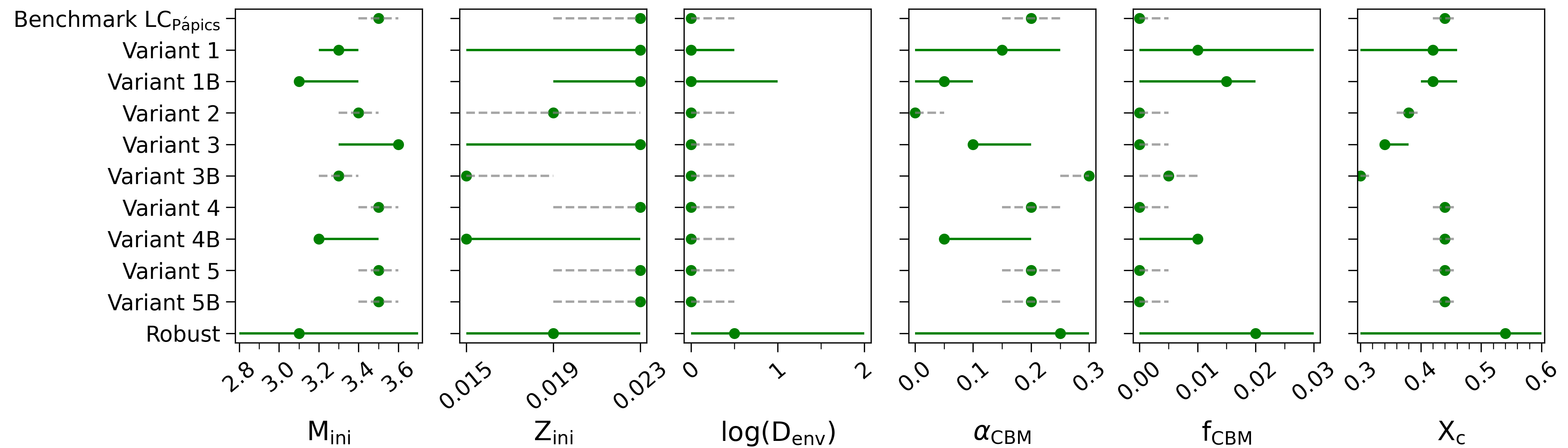}
	\caption{Parameters and 2$\sigma$ confidence intervals for benchmark, variant and robust period spacing patterns are shown as filled circles and solid lines, respectively. If a confidence interval calculation using Bayes’ theorem did not return any additional models, an upper limit estimate is taken to be the grid step size for a given parameter, which is indicated by the dashed grey lines.}
	\label{figure: manual parameters}
	\end{figure*}

	As discussed in section~\ref{subsubsection: subjective patterns}, we construct five variant patterns each containing only a single different frequency compared to the benchmark pattern, as shown in Fig.~\ref{figure: manual patterns}. For each of these variant patterns, we also create a counterpart in which the alternative frequency is simply omitted, which is referred to as the `B' pattern. This means that in the modelling of a `B' pattern, there is no enforcement of consecutive radial order for where the ambiguous frequency was omitted. An observed `B' pattern could potentially thus be best fitted by a theoretical pattern that contains none or multiple radial orders within the gap where the ambiguous frequency was omitted. This is different to the corresponding `A' pattern for which the best fitting theoretical pattern is forced to consist of 36 consecutive radial orders. Finally we include a `robust' pattern in which only unambiguous periods were included between $0.92 < P < 1.22$~d. For each variant pattern, we perform the same modelling setup as the benchmark model, with specific discussion for each case below. Summary figures for the modelling of variant patterns are shown in Figs~\ref{figure: manual modelling 1} and \ref{figure: manual modelling 2}, and derived parameter uncertainties using Bayes' theorem are shown in Fig.~\ref{figure: manual parameters}.
	
	\subsubsection{Variant patterns 1, 1B, 2 and 2B}
	
	In variant patterns 1 and 2, we select the middle and left-most frequencies from an apparent triplet of independent modes whereas in the benchmark pattern the right-most frequency was selected. This means that variant pattern 1 includes 0.8046355~d$^{-1}$ instead of 0.8029739~d$^{-1}$, and variant pattern 2 includes 0.8064031~d$^{-1}$ instead of 0.8029739~d$^{-1}$ as shown in Fig.~\ref{figure: manual patterns}. Since variant pattern 2 is made from the same triplet, there is no equivalent B pattern, since it would be the same as variant pattern 1B. Visually, the morphology of variant pattern 1 appears just as reasonable as the benchmark pattern: it is relatively smooth and periodic. Whereas the inclusion of the left-most frequency in the triplet in variant pattern 2 introduces a noticeable glitch between $1.24 < P < 1.26$~d.
	
	The summary figure for the modelling results of variant pattern 2 is shown in Fig.~\ref{figure: manual modelling 1}. The parameters and their uncertainties for variant patterns 1, 1B and 2 are shown in Fig.~\ref{figure: manual parameters}. In the case of variant patterns 1 and 2, the best models within the grid are no longer within the 3$\sigma$ spectroscopic error box, and neither is the best model from variant pattern 1B. However, this may be a chance circumstance owing to the sampling of the grid parameters given that the distribution of best models in the HR~diagram and individual parameters and uncertainties are similar to the benchmark model. In all cases, regardless of where the model with the lowest MD is located in the HR~diagram, the best model is always selected from within the 3$\sigma$ spectroscopic error box following \citet{Michielsen2021a}. The inclusion of an alternate frequency (variant patterns 1 and 2) or omitting the frequency (variant pattern 1B) yields parameters that are mostly consistent within 2$\sigma$ to the benchmark model, which is shown in Fig.~\ref{figure: manual parameters}. For example, the $X_{\rm c}$ value of variant pattern 2 is 0.38 compared to 0.44 in the benchmark model, but the 2$\sigma$ confidence interval for $X_{\rm c}$ in variant pattern 1 includes all the other values. This is also the case for $\alpha_{\rm CBM}$ and $f_{\rm CBM}$. Therefore, the inclusion of an alternate frequency and subsequent modelling of the different period spacing pattern at this specific location of a dip in the pattern of KIC~7760680 is sufficient to produce different confidence interval in some variant patterns, but that the parameters themselves are generally consistent.

	\subsubsection{Variant patterns 3 and 3B}
	
	In variant pattern 3, we alternatively select an adjacent frequency, 0.7335336~d$^{-1}$ instead of 0.7313646~d$^{-1}$, as illustrated in Fig.~\ref{figure: manual patterns}. Visually it is easy to see that this alternate frequency does not yield as satisfying a period spacing pattern, since it introduces a noticeable glitch between $1.36 < P < 1.38$~d. The summary figure for the modelling results of variant patterns 3 and 3B are shown in Fig.~\ref{figure: manual modelling 1}. The parameters and their uncertainties for variant patterns 3 and 3B are shown in Fig.~\ref{figure: manual parameters}. In the case of modelling variant pattern 3, the parameters and their uncertainties are generally consistent with the benchmark model at the 2$\sigma$ level. But a comparison of variant pattern 3B reveals that it is discrepant with the benchmark model, despite the solution being well constrained individually. Therefore, we conclude that it is preferable to include a potentially ambiguous frequency in an unbroken pattern of consecutive radial order g~modes rather than omit it entirely and fit two parts of a broken pattern.

	\subsubsection{Variant patterns 4 and 4B}
	
	In variant pattern 4, we alternatively select an adjacent frequency, 0.7609956~d$^{-1}$ instead of 0.7599243~d$^{-1}$ as illustrated in Fig.~\ref{figure: manual patterns}. The two frequencies involved in variant pattern 4 are special in this study because they both arise from an apparent doublet in the amplitude spectrum and are only separated by 1.6 times the Rayleigh resolution. Therefore, according to our iterative pre-whitening setup of keeping the highest amplitude frequency within 2.5 times the Rayleigh, only one of them was included in our list of frequencies in Table~\ref{table: multi-freq}. However, they statistically have the same amplitude: $0.06 \pm 0.01$~mmag.
	
	The summary figure for the modelling results of variant patterns 4 and 4B are shown in Fig.~\ref{figure: manual modelling 2}. The parameters and their uncertainties for variant patterns 4 and 4B are shown in Fig.~\ref{figure: manual parameters}. The exact same best model in the case of variant pattern 4 is returned as the benchmark model. This is not surprising since the alternative mode has only a very small (i.e. $\delta\,\nu = 0.001$~d$^{-1}$) frequency difference to the benchmark pattern. This implies that small frequency shifts and discarded spurious and unresolved frequencies according to the \citet{Loumos1978} do not strongly impact the modelling results. However, in the case of variant pattern 4B, in which neither frequency is included, much larger parameter uncertainties are returned. Hence, similar to the results based on modelling variant pattern 3B, we conclude it is preferable to not omit ambiguous frequencies as members of patterns. Moreover, this exercise is direct evidence of why high-precision space photometry of order 1~yr or longer is beneficial for forward asteroseismic modelling. Otherwise, with degraded frequency resolution from short light curves, incorrect or unresolved frequencies could be chosen.

	\subsubsection{Variant patterns 5 and 5B}
		
	Finally, in variant pattern 5 we intentionally select an `incorrect' frequency at the end of the pattern to test the modelling sensitivity to the inclusion of such a low-amplitude long-period mode. This means we include 0.6827688~d$^{-1}$ instead of 0.6846335~d$^{-1}$, as illustrated in Fig.~\ref{figure: manual patterns}. The summary figure for the modelling results of variant patterns 5 and 5B are shown in Fig.~\ref{figure: manual modelling 2}. The parameters and their uncertainties for variant patterns 5 and 5B are shown in Fig.~\ref{figure: manual parameters}. In this scenario, the impact on the modelling results is negligible; the same best model in the case of variant patterns 5 and 5B as the benchmark model are returned. This is interesting since such a frequency is the lowest radial order within a pattern for KIC~7760680, hence the tradeoff between increased probing power from additional radial orders and decreased precision because of its relatively large frequency uncertainty does not impact the modelling results. Or in other words, the inclusion of this particular frequency contributes to the goodness-of-fit metric in terms of the resultant MD value of the best model, but the model selection itself is not impacted by omitting this frequency entirely.

	\subsubsection{Robust pattern}

	Our investigation of testing the impact of different choices in building period spacing patterns from a list of significant pulsation frequencies has revealed that the same best model is not always returned, even if the alternative frequency only differs by 0.002~d$^{-1}$. In particular, a low envelope mixing value of $\log\,D_{\rm env} = 0.0$ is always returned for KIC~7760680, but $X_{\rm c}$ ranges between 0.30 and 0.54. On the other hand, the 2$\sigma$ confidence intervals in Fig.~\ref{figure: manual parameters} show that most solutions are generally consistent with each other. This may lead one to wonder how well constrained the solution space is if only the robust part of the period spacing pattern is used in modelling, that is only include periods in a pattern for which there is little-to-no ambiguity in which frequencies to include. In such a setup we restrict ourselves to only including high-amplitude modes with periods between $0.92 < P < 1.22$~d where the density of observed pulsation frequencies in the amplitude spectrum is much lower. For completeness, we also exclude the highest frequency from the pattern since its amplitude is quite low. This leaves us with 16 consecutive radial mode frequencies compared to the 36 in the benchmark and variant patterns (and 35 in the counterpart B patterns) discussed in this work. 
	
	The summary figure for the modelling results of the robust pattern is shown in Fig.~\ref{figure: manual modelling 2}. The parameters and their uncertainties for the robust pattern are shown in Fig.~\ref{figure: manual parameters}. The resultant distribution of the best models in the HR~diagram for the robust pattern is different to the benchmark model and the 2$\sigma$ confidence intervals are larger, in part because of the smaller number of observables in the fitting process: 16 mode periods instead of 36, but also because their corresponding probing powers are quite different. With more modes at lower radial orders in the benchmark pattern, the model parameters, such as mass, age, metallicity, are more precisely constrained because these modes are more sensitive to the convective core mass. Therefore, our comparison indicates that it is more beneficial to include additional modes at the expense of potentially larger observational uncertainties if one can maximise the radial order range and probing power of the g-mode period spacing pattern. The 2$\sigma$ confidence intervals for the parameters from the robust pattern shown in Fig.~\ref{figure: manual parameters} are relatively large and encompass all the best models of the benchmark and variant patterns discussed in this section. Therefore, we conclude that such a modelling strategy is the most accurate but not as precise.


\section{Conclusions}
\label{section: conclusions}

In this work, we use the SPB star KIC~7660680 observed by the \Kepler space telescope to test the impact of different choices and setups in the extraction of g-mode frequencies from space photometry on the determination of the best model parameters within a forward asteroseismic modelling framework. We use three representative light curves, denoted as \sap{}, \pdc{} and \pap{}, with the former two being derived from the MAST SAP and PDC-SAP light curves, respectively, and the third being the light curve extracted by \citet{Papics2015} in their discovery paper of KIC~7760680. We demonstrate that the \pap{} light curve is superior for a consistent frequency analysis. Hence we conclude that forward asteroseismic modelling studies should extract an optimised light curve, since the results of a subsequent frequency analysis become divergent among the three \sap{}, \pdc{} and \pap{} light curves. 

We also investigate the impact of including a multi-cosinusoid non-linear least-squares fit during iterative pre-whitening, and conclude that a significantly different frequency list is produced. Studies that use the Fourier frequencies as input to forward asteroseismic modelling and estimate observational uncertainties based on the idealistic formulae of \citet{Montgomery1999} are underestimating the true frequency uncertainty. The significant differences in the resultant frequency list when a multi-cosinusoid non-linear least-squares fit is performed are understandable since pulsation mode frequencies are not completely independent. Such a fit and the resultant correlated uncertainties incorporate the covariance of all the frequencies, amplitudes and phases, with frequencies and phases being highly correlated parameters. Therefore we conclude that such a step is necessary to produce an accurate and self-consistent list of pulsation mode frequencies.

Based on a set of simulated light curves emulating a 4-yr \Kepler light curve, we revisit the choice of significance criterion in iterative pre-whitening. Our simulations demonstrate S/N~$\geq 4.6$ corresponds to a false alarm probability of 1\% that an extracted frequency corresponds to a white noise peak. \citet{Zong2016a} used similar simulations to show that the amplitude S/N~$\geq 4$ criterion of \citet{Breger1993b} is too low for {\it Kepler}-like light curves. Hence we advocate caution in interpreting signals obtained from iterative pre-whitening below S/N~$< 4.6$ with \Kepler light curves. Furthermore, \citet{Baran2015b} advise S/N~$\geq 5$ in the case of light curves from the K2 mission given their increased likelihood of containing instrumental systematics. Such a conservative significance criterion is also applicable to analysis of light curves from the ongoing TESS mission (see e.g. \citealt{Burssens2020a, Baran2021c}).

To assess the propagation of observational uncertainties in forward asteroseismic modelling, we create three sets of 100 perturbed period spacing patterns for each of the \sap{}, \pdc{} and \pap{} light curves. Each set of perturbed patterns took the original pattern of \citet{Papics2015} as input but it is optimised using a multi-cosinusoid non-linear least-squares fit to the respective light curve. Each of the three \sap{}, \pdc{} and \pap{} solutions are randomly perturbed within their 1$\sigma$ frequency uncertainties to create the 300 perturbed patterns. Following \citet{Michielsen2021a}, we use 4-, 5- and 6-parameter modelling setups to determine the best model for each perturbed pattern. We find that the 4-parameter setup is insensitive to the observational uncertainties, but a large spread in best models and inferred core masses are returned for the 5-parameter setup. The majority of the returned best models in the 5-parameter setup are different to the best model found by \citet{Michielsen2021a} based on the pattern of \cite{Papics2015}. However, in the 6-parameter setup, the vast majority of best models are consistent. Hence we conclude that the 6-parameter modelling setup is more robust against random observational uncertainties that may be introduced from an imperfect light curve and pulsation mode frequency extraction through iterative pre-whitening.

Finally, we investigate the impact of potentially ambiguous pulsation frequencies as members of period spacing patterns. Using five examples, we build variant period spacing patterns in which a single frequency is replaced by an adjacent one. We also create a corresponding set of `B' counterpart patterns in which the alternative frequency is removed entirely to test if it is preferable to simply remove an ambiguous frequency within a pattern. The resultant best models for each of the five variant patterns and their `B' counterparts demonstrate that the inclusion of a potentially ambiguous frequency can impact the confidence intervals of the resultant model parameters, but this depends on the frequency (i.e. radial order) of the specific pulsation mode. Moreover, it is generally preferable to include a potentially ambiguous frequency instead of omitting them entirely from the pattern. Our results show that some pulsation modes have more weight than others in determining the best model based on where they are located within a pattern. The inclusion of specific frequencies introduce perturbations to the morphology of a period spacing pattern, which if are sufficiently close to a dip in the pattern caused by the chemical gradient left behind from the shrinking core during the main sequence (see e.g. \citealt{Miglio2008a}) yields a significantly different best model.

Interestingly, in the case of the benchmark pattern and variant patterns 1 and 2, none of the three possible frequencies chosen from an apparent triplet produce a period spacing pattern which is well fitted by our models. We posit that there is missing physics, such as the presence of a magnetic field in the deep interior of KIC~7760680 (see e.g. \citealt{Prat2019a, Prat2020a, VanBeeck2020a}), which is causing a potential frequency perturbation at the location of the frequency triplet between $0.80 < \nu < 0.81$~d$^{-1}$. In that sense, the theoretical uncertainties associated with incomplete physics within the models remain a dominant source of uncertainty in forward asteroseismic modelling of main-sequence stars.

Our results are based on the in-depth analysis of a single star, but they are extendable to other coherent g-mode pulsators since KIC~7760680 is an excellent representative of SPB stars. The \Kepler sample of SPB stars span a diverse range of masses, ages, and rotation rates (see e.g. \citealt{Szewczuk2021a, Pedersen2021a}), but all show interesting features and structure in their observed g-mode period spacing patterns, which models cannot fully explain. Our sensitivity study has demonstrated that the formal precision on pulsation frequencies obtained using exquisite \Kepler space photometry whilst taking into account various data reduction steps are small but non-negligible compared to the larger theoretical uncertainties, and can be used to effectively fine-tune the physics of current theoretical models. In conclusion, even though theoretical uncertainties associated with incomplete physics within the models dominate in forward asteroseismic modelling, a more systematic treatment of observational uncertainties allows us to better constrain the interior physics and determine more precise and accurate (convective core) masses, ages, and interior rotation and mixing parameters in excellent g-mode pulsators such as KIC~7760680.


\begin{acknowledgements}
The authors are grateful to the asteroseismology group at KU Leuven for useful discussions and the referee for constructive feedback. The research leading to these results has received funding from the Research Foundation Flanders (FWO) by means of a senior postdoctoral fellowship to DMB (grant agreement No.~1286521N), and a PhD scholarship to MM (grant agreement No.~11F7120N).

The authors thank the \Kepler science team for the excellent data. The \Kepler data presented in this paper were obtained from the Mikulski Archive for Space Telescopes (MAST) at the Space Telescope Science Institute (STScI), which is operated by the Association of Universities for Research in Astronomy, Inc., under NASA contract NAS5-26555. Support to MAST for these data is provided by the NASA Office of Space Science via grant NAG5-7584 and by other grants and contracts. Funding for the Kepler/K2 mission was provided by NASA’s Science Mission Directorate. This research has made use of the SIMBAD database, operated at CDS, Strasbourg, France; the SAO/NASA Astrophysics Data System; and the VizieR catalog access tool, CDS, Strasbourg, France. The computational resources and services used in this work were provided by the VSC (Flemish Supercomputer Center), funded by the Research Foundation Flanders (FWO) and the Flemish Government, department EWI.

The authors are grateful to the {\sc MESA} \citep{Paxton2019} and {\sc GYRE} \citep{Townsend2013b} developers, in particular B. Paxton and R. H. D. Townsend, for continually supporting the development of state-of-the-art and open-source tools for modelling pulsating stars. This research has made use of the \texttt{PYTHON} library for publication quality graphics (\texttt{MATPLOTLIB}; \citealt{Matplotlib_2007}), \texttt{Seaborn} \citep{seaborn_2021}, \texttt{Numpy} \citep{Numpy_2006, Numpy_2011, Numpy_2020} and \texttt{Pandas} \citep{Pandas_2010}.

\end{acknowledgements}


\bibliographystyle{aa}
\bibliography{/Users/dominic/Documents/RESEARCH/Bibliography/master_bib}


\onecolumn

\begin{appendix}

\section{Extended data tables}
\label{section: appendix: tables}

The multi-cosinusoid non-linear least-squares solution for the \pap{} light curve of KIC~7760680 is provided in Table~\ref{table: multi-freq}.

\longtab{
\begin{longtable}{c r c r c}
\caption{\label{table: multi-freq} Multi-cosinusoid non-linear least-squares solution for the \pap{} light curve of KIC~7760680 for frequencies between $0.2 \leq \nu \leq 1.5$~d$^{-1}$ using the multi-cosinusoid regression model (cf. Eqn.~\ref{equation: cosine}). The zero point of the time scale is $t_0 = 2455688.770$~BJD. The final column shows the absolute frequency difference between the multi-cosinusoid and single-frequency solutions (cf. Table~\ref{table: single-freq}), and is expressed in terms of $\sigma$ of the multi-cosinusoid solution.} \\
\hline\hline
\multicolumn{1}{c}{Freq. label} & \multicolumn{1}{c}{Frequency} & \multicolumn{1}{c}{Amplitude} & \multicolumn{1}{c}{Phase} & \multicolumn{1}{c}{Freq. Diff.} \\
\multicolumn{1}{c}{} & \multicolumn{1}{c}{} & \multicolumn{1}{c}{$\pm 0.005$} & \multicolumn{1}{c}{} & \multicolumn{1}{c}{} \\
\multicolumn{1}{c}{} & \multicolumn{1}{c}{(d$^{-1}$)} & \multicolumn{1}{c}{(mmag)} & \multicolumn{1}{c}{(rad)} & \multicolumn{1}{c}{($\sigma_{\nu}$)} \\
\hline
\endfirsthead
\caption{\it continued.}\\
\hline\hline
\multicolumn{1}{c}{Freq. label} & \multicolumn{1}{c}{Frequency} & \multicolumn{1}{c}{Amplitude} & \multicolumn{1}{c}{Phase} & \multicolumn{1}{c}{Freq. Diff.} \\
\multicolumn{1}{c}{} & \multicolumn{1}{c}{} & \multicolumn{1}{c}{$\pm 0.005$} & \multicolumn{1}{c}{} & \multicolumn{1}{c}{} \\
\multicolumn{1}{c}{} & \multicolumn{1}{c}{(d$^{-1}$)} & \multicolumn{1}{c}{(mmag)} & \multicolumn{1}{c}{(rad)} & \multicolumn{1}{c}{($\sigma_{\nu}$)} \\
\hline
\endhead
\hline
\endfoot

$1$		&	$0.8628752 \pm 0.0000002$	&	$10.704$		&	$0.4996 \pm 0.0004$	&	$15.7$	\\
$2$		&	$0.8385259 \pm 0.0000003$	&	$5.482$		&	$-2.2699 \pm 0.0008$	&	$4.6$	\\
$3$		&	$0.8948482 \pm 0.0000005$	&	$3.706$		&	$2.2675 \pm 0.0012$	&	$1.1$	\\
$4$		&	$0.8143109 \pm 0.0000009$	&	$1.889$		&	$-1.1029 \pm 0.0024$	&	$8.3$	\\
$5$		&	$0.9921849 \pm 0.0000011$	&	$1.512$		&	$1.6427 \pm 0.0030$	&	$4.0$	\\
$6$		&	$0.9133844 \pm 0.0000017$	&	$1.028$		&	$1.8397 \pm 0.0044$	&	$2.2$	\\
$7$		&	$0.9318244 \pm 0.0000025$	&	$0.672$		&	$-0.6662 \pm 0.0067$	&	$0.1$	\\
$8$		&	$0.9678453 \pm 0.0000029$	&	$0.586$		&	$-1.3360 \pm 0.0077$	&	$1.6$	\\
$9$		&	$0.7942531 \pm 0.0000034$	&	$0.510$		&	$0.5646 \pm 0.0089$	&	$1.8$	\\
$12$		&	$0.7457526 \pm 0.0000045$	&	$0.385$		&	$2.0895 \pm 0.0117$	&	$2.2$	\\
$15$		&	$0.8505494 \pm 0.0000058$	&	$0.294$		&	$-1.2277 \pm 0.0152$	&	$8.0$	\\
$16$		&	$0.8263110 \pm 0.0000064$	&	$0.269$		&	$-2.2593 \pm 0.0167$	&	$0.8$	\\
$17$		&	$0.8772515 \pm 0.0000070$	&	$0.247$		&	$1.7804 \pm 0.0183$	&	$6.0$	\\
$19$		&	$1.0186482 \pm 0.0000071$	&	$0.238$		&	$0.2539 \pm 0.0188$	&	$0.4$	\\
$20$		&	$0.9697119 \pm 0.0000107$	&	$0.160$		&	$3.0847 \pm 0.0283$	&	$0.2$	\\
$21$		&	$0.8161805 \pm 0.0000103$	&	$0.174$		&	$-0.8090 \pm 0.0262$	&	$0.3$	\\
$24$		&	$0.7531799 \pm 0.0000130$	&	$0.130$		&	$1.7564 \pm 0.0341$	&	$0.8$	\\
$27$		&	$0.8029739 \pm 0.0000133$	&	$0.140$		&	$-1.5464 \pm 0.0322$	&	$3.0$	\\
$28$		&	$0.8046355 \pm 0.0000145$	&	$0.135$		&	$1.3832 \pm 0.0340$	&	$1.1$	\\
$29$		&	$0.7090951 \pm 0.0000139$	&	$0.124$		&	$-3.0163 \pm 0.0365$	&	$2.4$	\\
$30$		&	$0.8308642 \pm 0.0000136$	&	$0.127$		&	$-1.6133 \pm 0.0357$	&	$1.6$	\\
$34$		&	$0.7335336 \pm 0.0000169$	&	$0.103$		&	$-0.5023 \pm 0.0437$	&	$4.1$	\\
$38$		&	$0.6736307 \pm 0.0000193$	&	$0.089$		&	$-2.6274 \pm 0.0507$	&	$1.0$	\\
$41$		&	$0.7137540 \pm 0.0000206$	&	$0.084$		&	$2.3207 \pm 0.0539$	&	$1.3$	\\
$44$		&	$0.9485893 \pm 0.0000210$	&	$0.081$		&	$3.0759 \pm 0.0555$	&	$0.2$	\\
$47$		&	$0.8064031 \pm 0.0000226$	&	$0.081$		&	$1.7194 \pm 0.0558$	&	$7.6$	\\
$52$		&	$0.9454690 \pm 0.0000265$	&	$0.064$		&	$-1.7554 \pm 0.0699$	&	$1.3$	\\
$55$		&	$0.6827688 \pm 0.0000264$	&	$0.067$		&	$1.1964 \pm 0.0683$	&	$0.4$	\\
$56$		&	$0.7839893 \pm 0.0000287$	&	$0.061$		&	$-2.6545 \pm 0.0748$	&	$1.8$	\\
$57$		&	$0.7194313 \pm 0.0000216$	&	$0.080$		&	$-2.1618 \pm 0.0567$	&	$3.3$	\\
$59$		&	$0.7609956 \pm 0.0000288$	&	$0.061$		&	$2.1391 \pm 0.0755$	&	$6.5$	\\
$68$		&	$0.8870805 \pm 0.0000359$	&	$0.048$		&	$-2.9368 \pm 0.0940$	&	$1.7$	\\
$70$		&	$0.6650084 \pm 0.0000305$	&	$0.057$		&	$0.4682 \pm 0.0801$	&	$0.7$	\\
$78$		&	$0.7384767 \pm 0.0000370$	&	$0.048$		&	$-1.7648 \pm 0.0968$	&	$5.7$	\\
$81$		&	$1.0747771 \pm 0.0000350$	&	$0.049$		&	$-2.8024 \pm 0.0925$	&	$2.0$	\\
$82$		&	$1.0452435 \pm 0.0000352$	&	$0.049$		&	$0.7727 \pm 0.0931$	&	$0.9$	\\
$89$		&	$1.1093061 \pm 0.0000373$	&	$0.045$		&	$2.1816 \pm 0.0987$	&	$0.2$	\\
$90$		&	$0.7672889 \pm 0.0000434$	&	$0.042$		&	$-2.0006 \pm 0.1138$	&	$2.9$	\\
$101$	&	$0.9189108 \pm 0.0000484$	&	$0.035$		&	$-1.3541 \pm 0.1277$	&	$5.7$	\\
$105$	&	$0.7034648 \pm 0.0000433$	&	$0.040$		&	$-2.3213 \pm 0.1137$	&	$1.2$	\\
$108$	&	$0.6978087 \pm 0.0000730$	&	$0.023$		&	$2.3761 \pm 0.1918$	&	$0.4$	\\
$111$	&	$0.8701929 \pm 0.0000402$	&	$0.036$		&	$-1.3117 \pm 0.1056$	&	$8.8$	\\
$114$	&	$0.7756747 \pm 0.0000432$	&	$0.041$		&	$-3.1501 \pm 0.1127$	&	$3.6$	\\
$116$	&	$0.7169843 \pm 0.0000542$	&	$0.032$		&	$-0.3889 \pm 0.1413$	&	$0.1$	\\
$121$	&	$0.7246552 \pm 0.0000488$	&	$0.035$		&	$0.3931 \pm 0.1278$	&	$6.8$	\\
$122$	&	$0.7313646 \pm 0.0000438$	&	$0.040$		&	$1.4010 \pm 0.1150$	&	$1.6$	\\
$131$	&	$0.6342698 \pm 0.0000585$	&	$0.029$		&	$-0.6788 \pm 0.1544$	&	$1.1$	\\
$134$	&	$0.7880626 \pm 0.0000949$	&	$0.023$		&	$2.7933 \pm 0.2488$	&	$0.7$	\\
$136$	&	$0.8812344 \pm 0.0000382$	&	$0.047$		&	$0.0407 \pm 0.1000$	&	$0.1$	\\
$147$	&	$0.7795457 \pm 0.0000441$	&	$0.041$		&	$0.4999 \pm 0.1142$	&	$0.6$	\\
$148$	&	$0.6846335 \pm 0.0000463$	&	$0.037$		&	$0.5175 \pm 0.1209$	&	$0.8$	\\
$149$	&	$0.6502099 \pm 0.0000733$	&	$0.023$		&	$2.6788 \pm 0.1935$	&	$1.1$	\\
$152$	&	$0.8554639 \pm 0.0000361$	&	$0.030$		&	$2.1391 \pm 0.0944$	&	$11.9$	\\
$154$	&	$0.6884050 \pm 0.0000604$	&	$0.028$		&	$-1.0619 \pm 0.1587$	&	$0.8$	\\
$156$	&	$0.7363277 \pm 0.0000540$	&	$0.031$		&	$2.4725 \pm 0.1397$	&	$3.2$	\\
$157$	&	$0.7695100 \pm 0.0001146$	&	$0.017$		&	$-1.1837 \pm 0.2988$	&	$0.5$	\\
$161$	&	$0.8183324 \pm 0.0000450$	&	$0.039$		&	$-3.0039 \pm 0.1176$	&	$1.2$	\\
$163$	&	$0.9376644 \pm 0.0000443$	&	$0.039$		&	$-2.1056 \pm 0.1166$	&	$1.7$	\\
$168$	&	$0.8442741 \pm 0.0000458$	&	$0.037$		&	$1.0457 \pm 0.1193$	&	$1.4$	\\
$175$	&	$0.6605905 \pm 0.0000913$	&	$0.019$		&	$-1.4798 \pm 0.2406$	&	$2.7$	\\
$185$	&	$0.7902605 \pm 0.0000451$	&	$0.037$		&	$2.1261 \pm 0.1184$	&	$8.2$	\\

\end{longtable}
}


\section{Variant pattern modelling results}
\label{section: appendix: figures}

\begin{figure*}
\centering
\includegraphics[width=0.49\columnwidth]{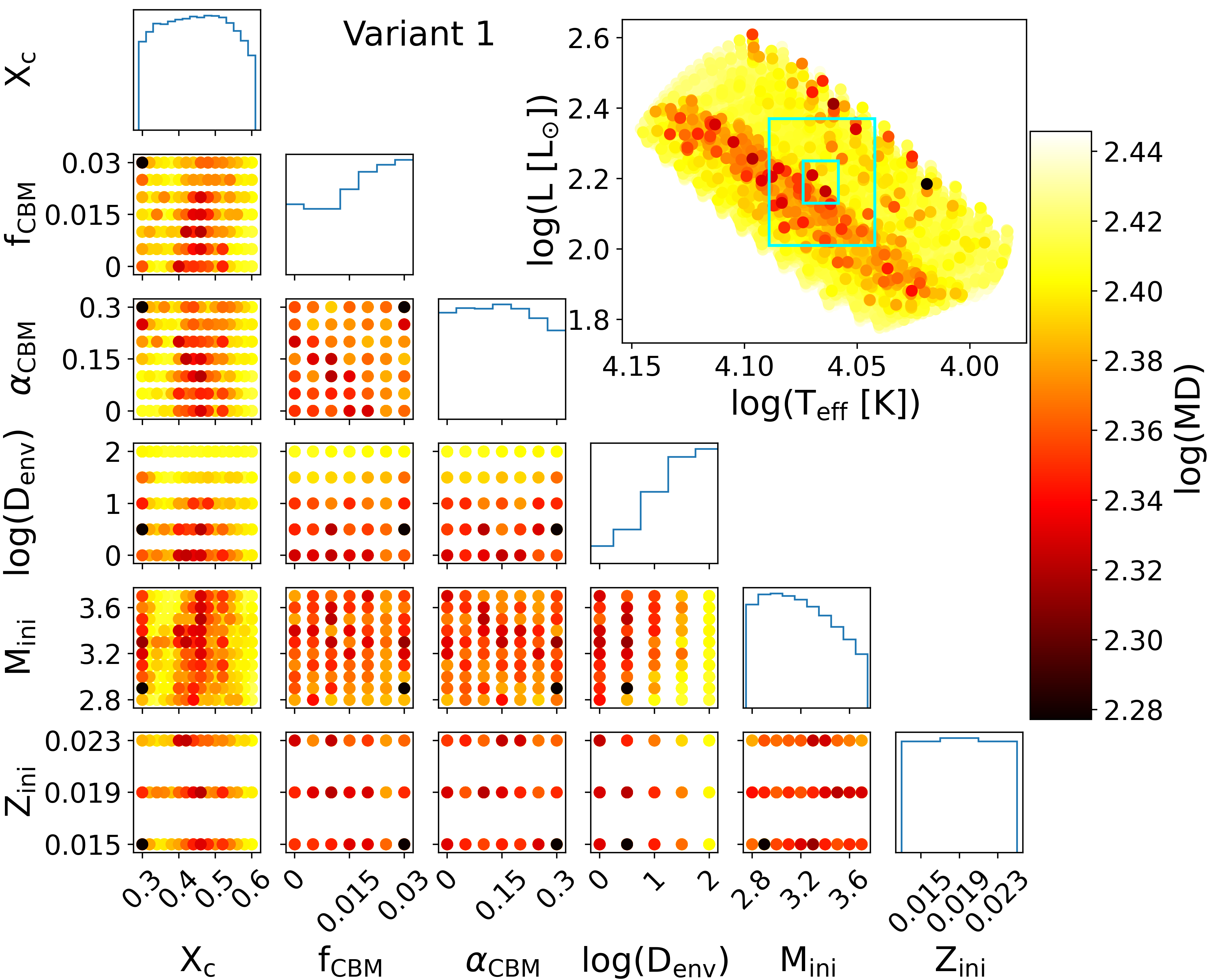}  \vspace{1cm}
\includegraphics[width=0.49\columnwidth]{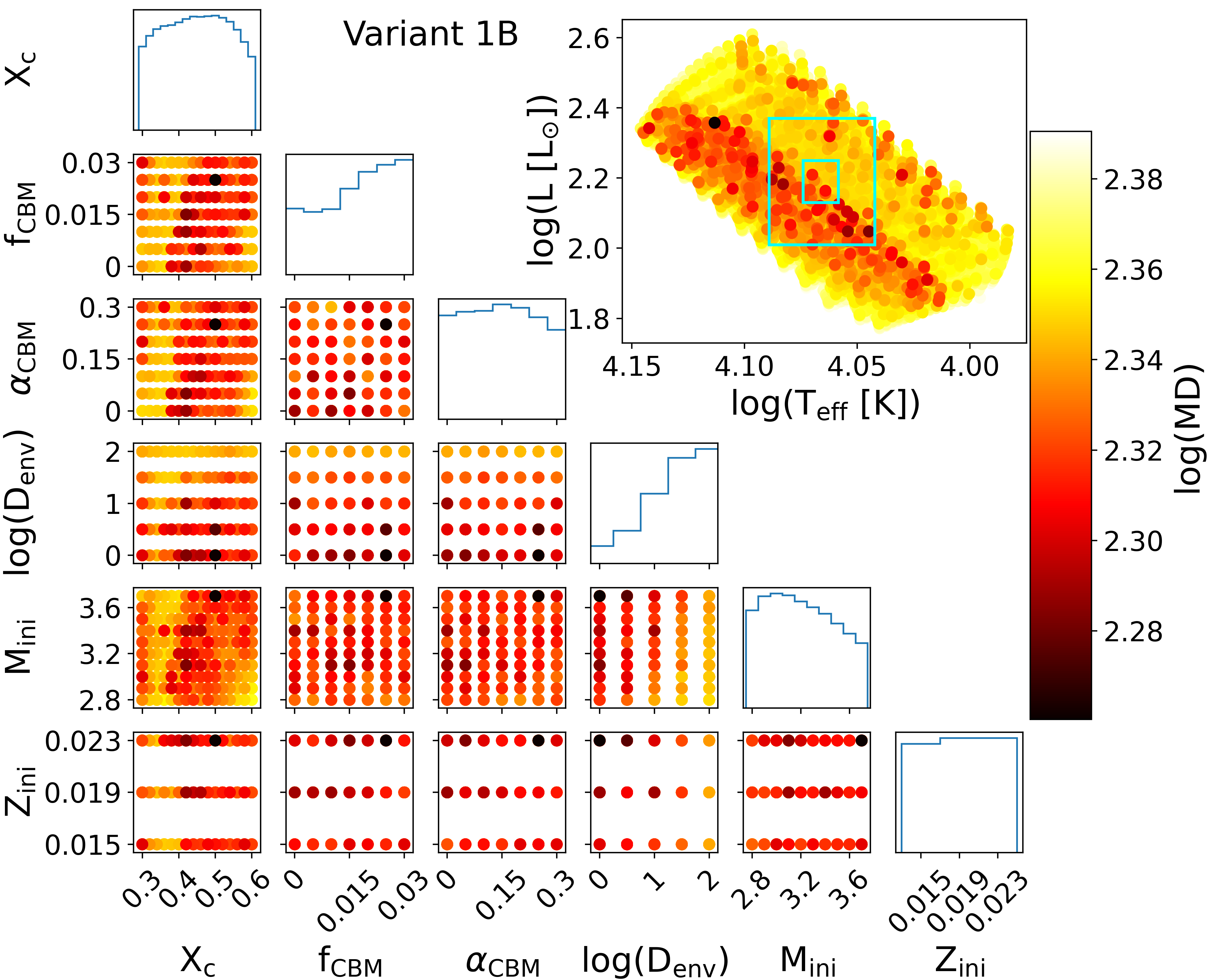}
\includegraphics[width=0.49\columnwidth]{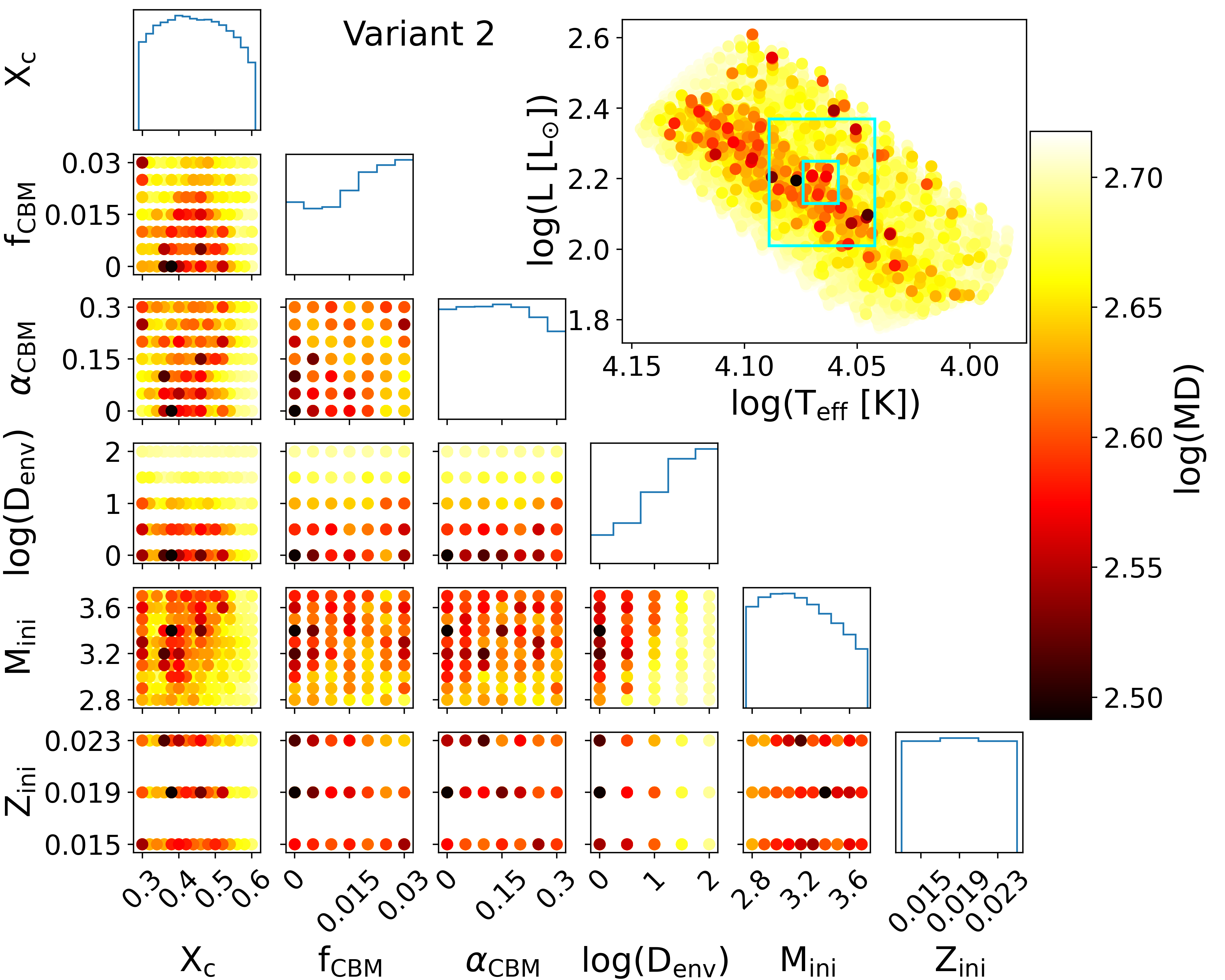} \vspace{1cm} \\
\includegraphics[width=0.49\columnwidth]{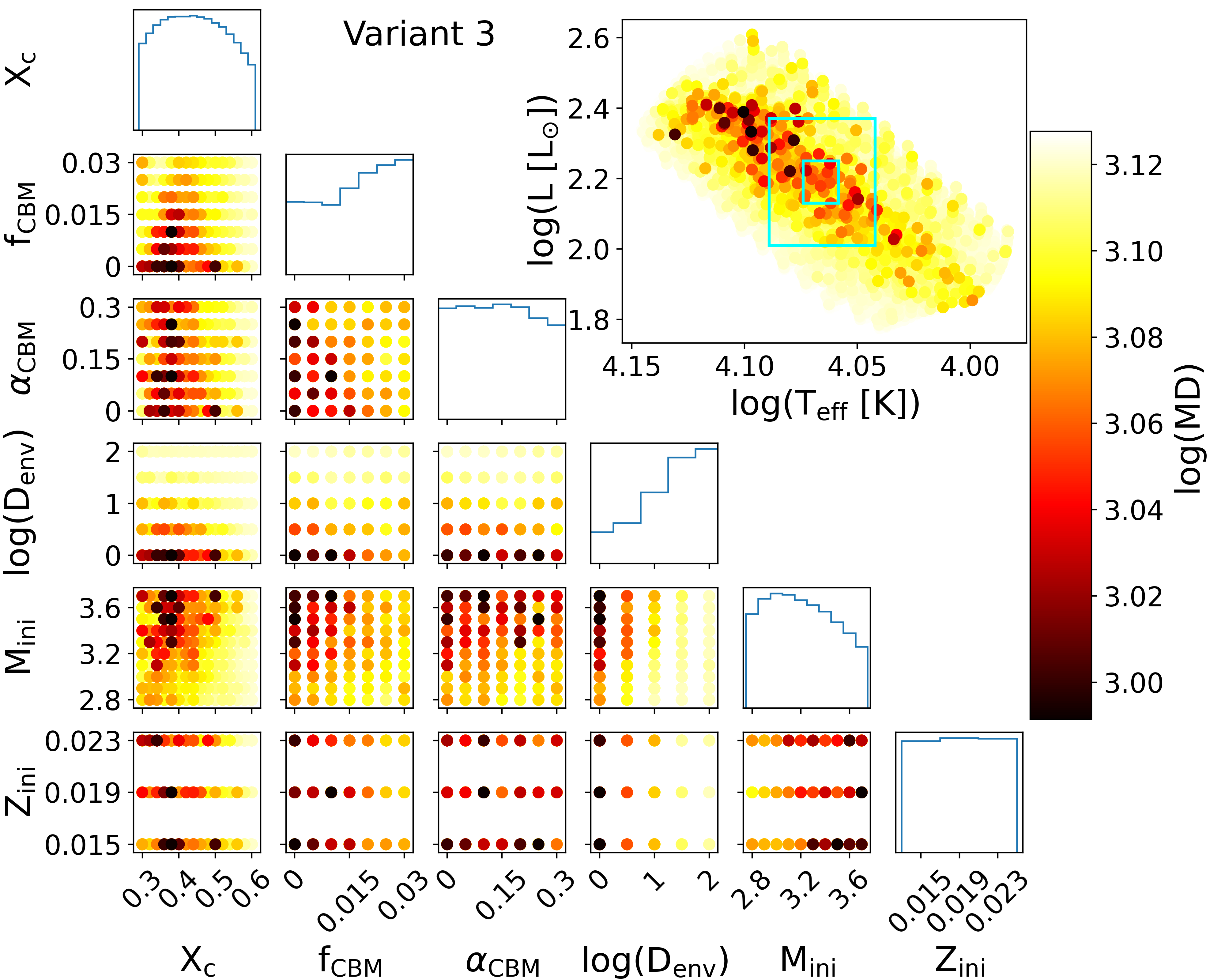} 
\includegraphics[width=0.49\columnwidth]{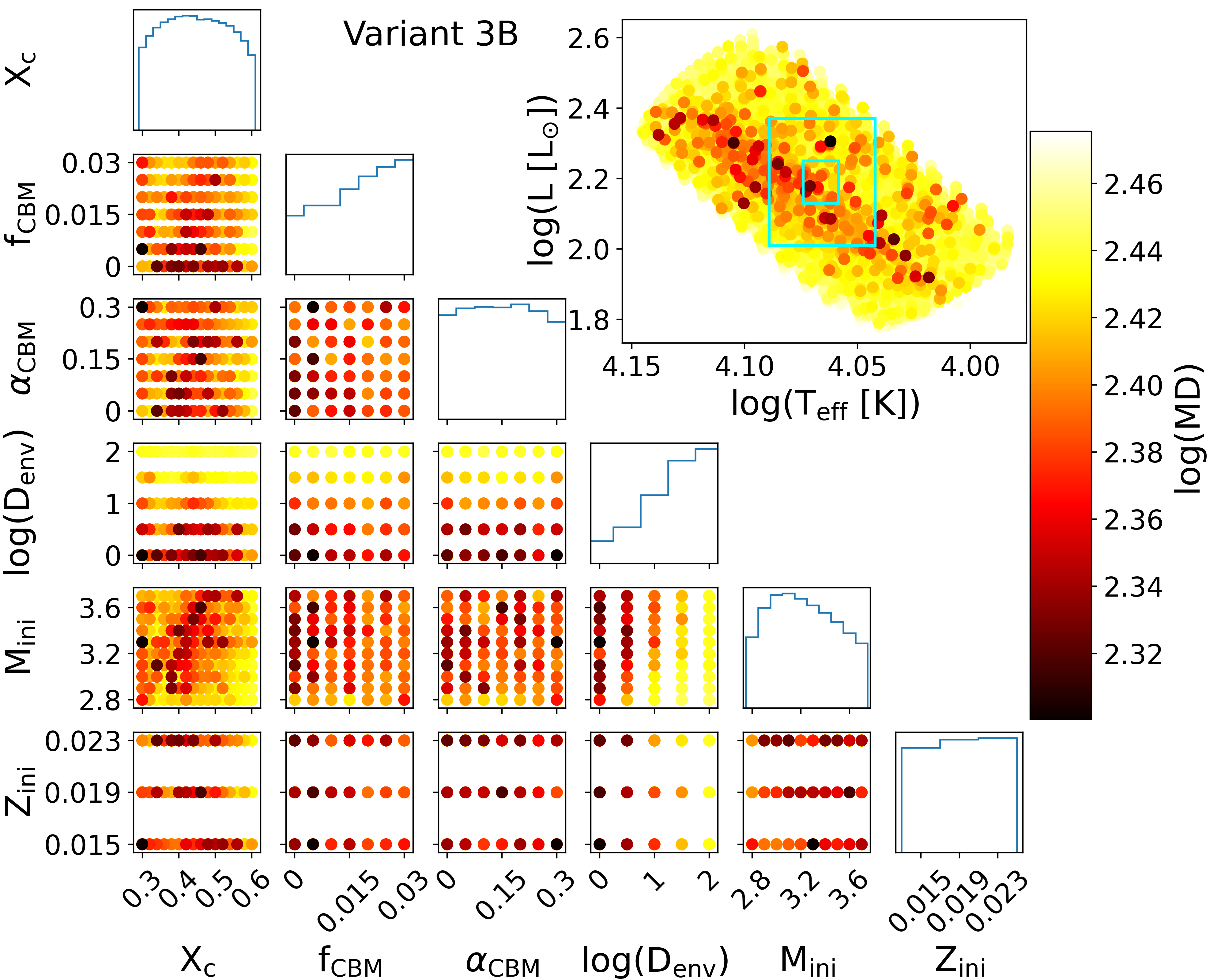}
\caption{Summary figures for the results from modelling variant period spacing patterns 1, 1B, 2, 3 and 3B.}
\label{figure: manual modelling 1}
\end{figure*}

\begin{figure*}
\centering
\includegraphics[width=0.49\columnwidth]{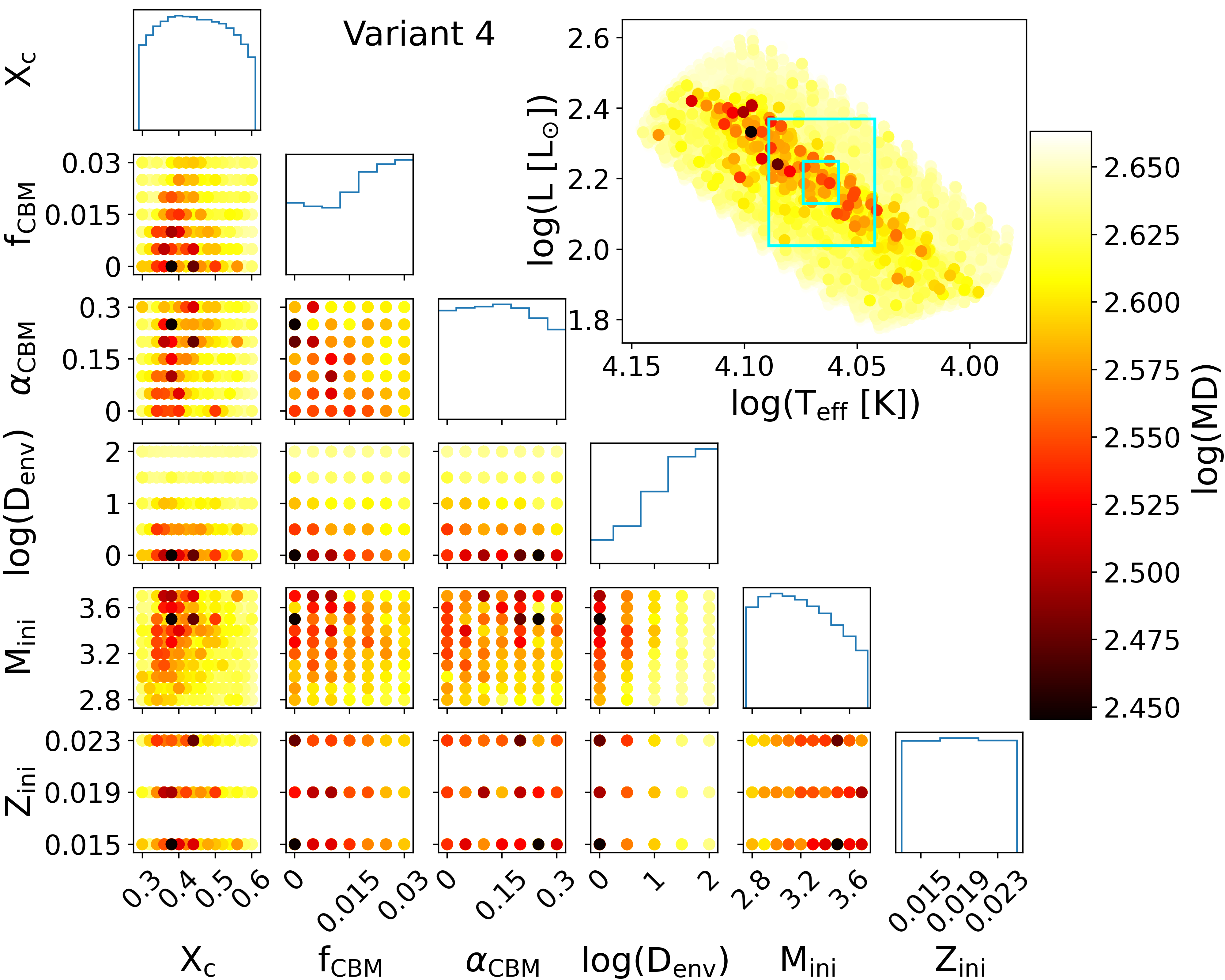}  \vspace{1cm}
\includegraphics[width=0.49\columnwidth]{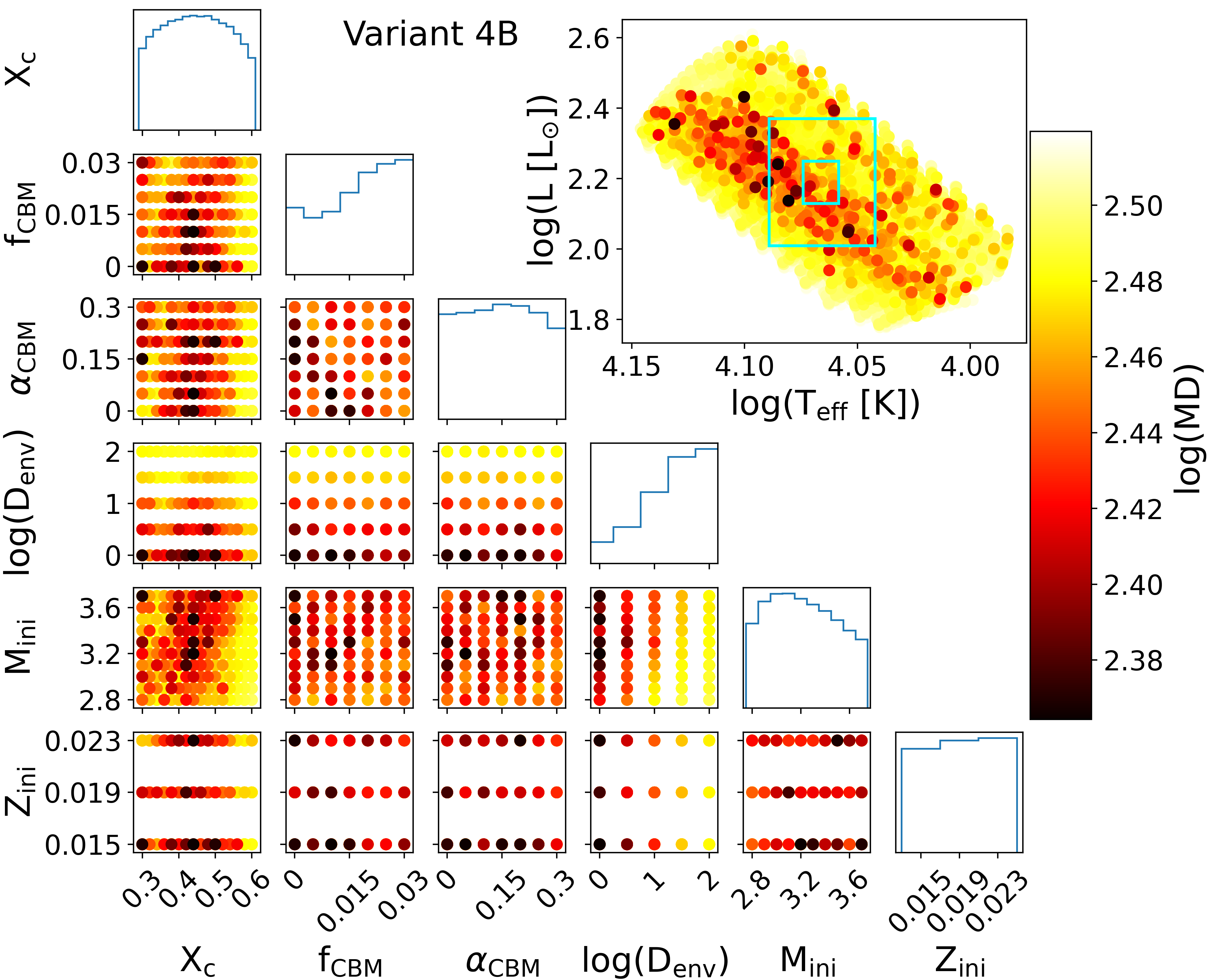}
\includegraphics[width=0.49\columnwidth]{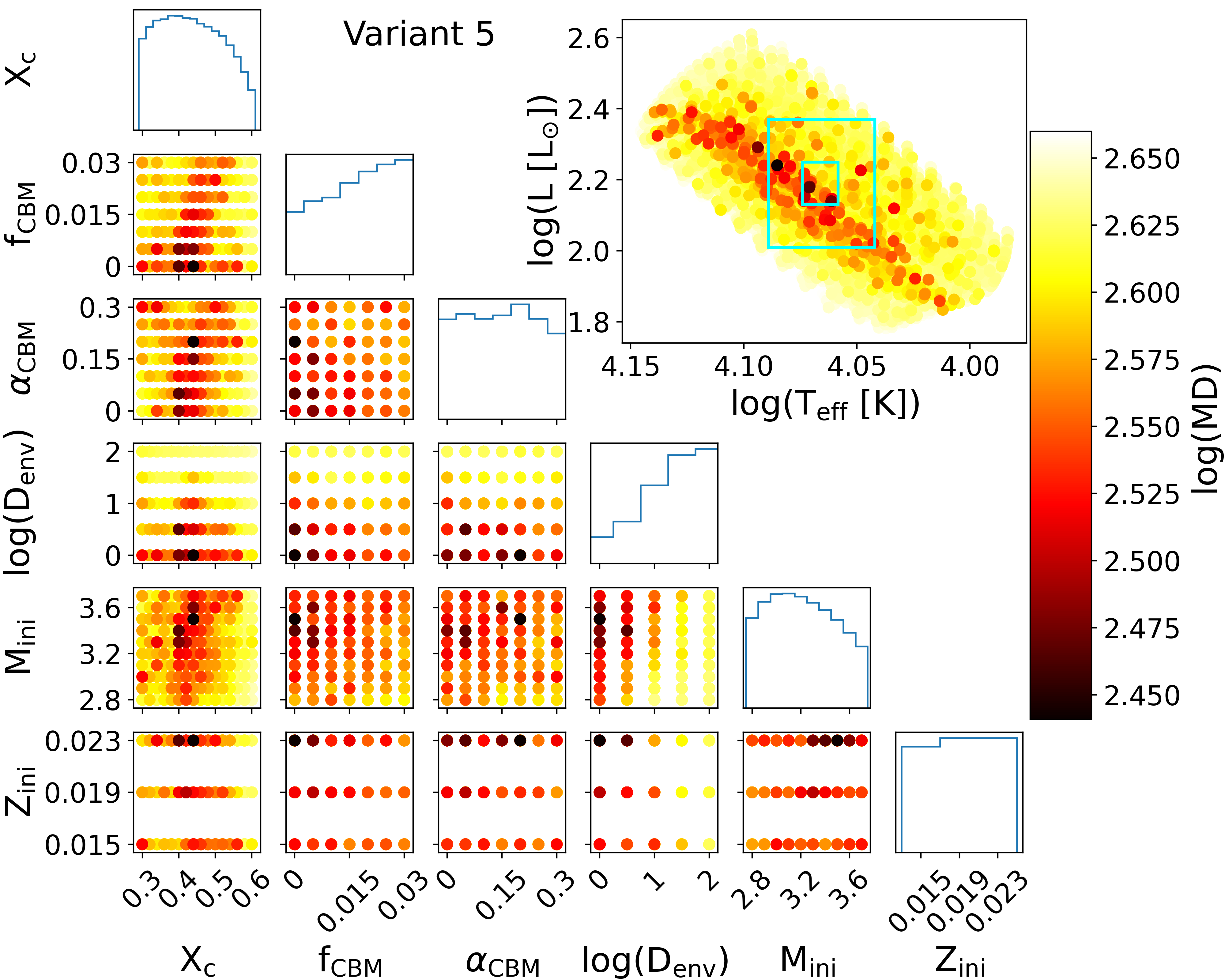}  \vspace{1cm}
\includegraphics[width=0.49\columnwidth]{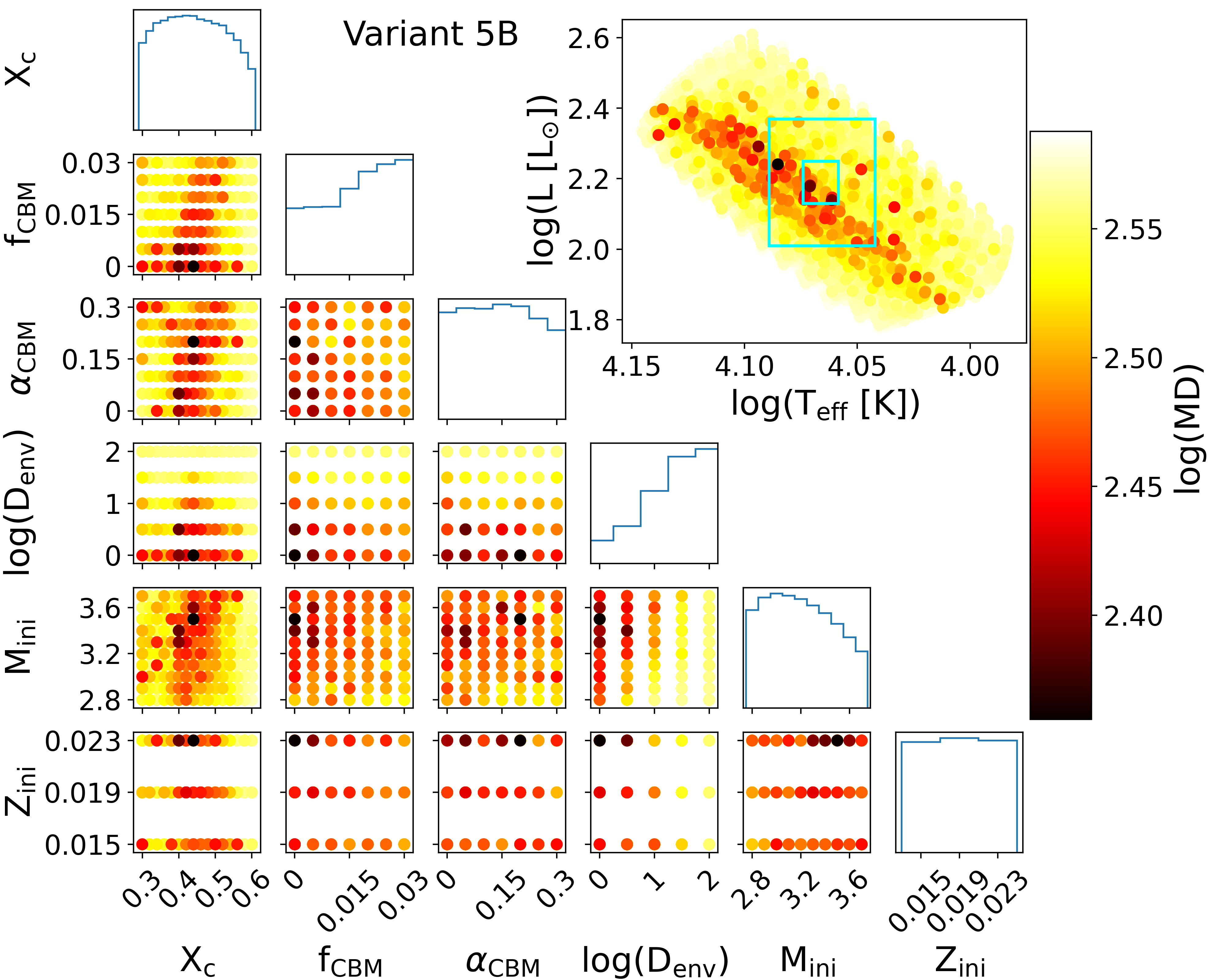}	
\includegraphics[width=0.49\columnwidth]{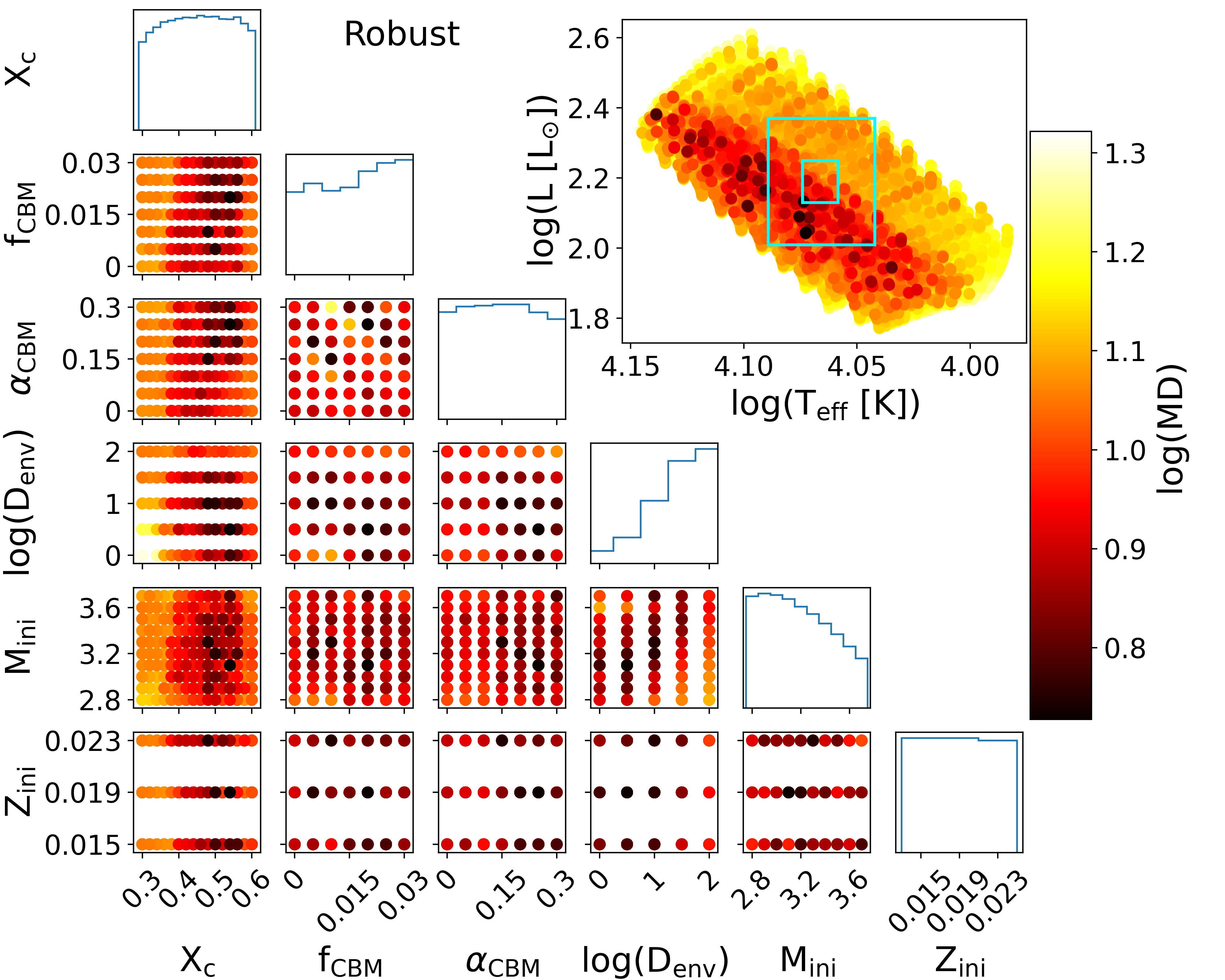}
\caption{Summary figures for the results from modelling variant period spacing patterns 4, 4B, 5, 5B, and the robust pattern.}
\label{figure: manual modelling 2}
\end{figure*}

The summary figures for modelling the variant period spacing patterns are shown in Figs~\ref{figure: manual modelling 1} and \ref{figure: manual modelling 2}.

\end{appendix}


\end{document}